\documentclass[aps,prl,reprint,superscriptaddress,amsmath,amssymb]{revtex4-2}
\usepackage{graphicx}
\usepackage{dcolumn}
\usepackage{bm}
\usepackage{gensymb}
\usepackage{hyperref}
\usepackage{color}
\newcommand{\bvec}[1]{\boldsymbol{#1}}

\begin{document}
\title{Moiré Engineering of Nonsymmorphic Symmetries and Hourglass Superconductors }

\author{Yifan Gao}
\altaffiliation{These authors contributed equally.}
\affiliation{Department of Physics, Southern University of Science and Technology, Shenzhen, Guangdong 518055, China}

\author{Ammon Fischer}
\altaffiliation{These authors contributed equally.}
\affiliation{Institut f\"ur Theorie der Statistischen Physik, RWTH Aachen University and JARA-Fundamentals of Future Information Technology, 52056 Aachen, Germany}

\author{Lennart Klebl}
\affiliation{Institut f\"ur Theorie der Statistischen Physik, RWTH Aachen University and JARA-Fundamentals of Future Information Technology, 52056 Aachen, Germany}

\author{Martin Claassen}
\email[Email: ]{claassen@sas.upenn.edu}
\affiliation{Department of Physics and Astronomy, University of Pennsylvania, Philadelphia, PA 19104}

\author{Angel Rubio}
\email[Email: ]{Angel.Rubio@mpsd.mpg.de}
\affiliation{Max Planck Institute for the Structure and Dynamics of Matter, Center for Free Electron Laser Science, 22761 Hamburg, Germany}
\affiliation{Center for Computational Quantum Physics, Simons Foundation Flatiron Institute, New York, NY 10010 USA}

\author{Li Huang}
\email[Email: ]{huangl@sustech.edu.cn}
\affiliation{Department of Physics, Southern University of Science and Technology, Shenzhen, Guangdong 518055, China}

\author{Dante Kennes}
\email[Email: ]{Dante.Kennes@rwth-aachen.de}
\affiliation{Institut f\"ur Theorie der Statistischen Physik, RWTH Aachen University and JARA-Fundamentals of Future Information Technology, 52056 Aachen, Germany}
\affiliation{Max Planck Institute for the Structure and Dynamics of Matter, Center for Free Electron Laser Science, 22761 Hamburg, Germany}

\author{Lede Xian}
\email[Email: ]{xianlede@sslab.org.cn}
\affiliation{Songshan Lake Materials Laboratory, 523808 Dongguan, Guangdong, China}
\affiliation{Max Planck Institute for the Structure and Dynamics of Matter, Center for Free Electron Laser Science, 22761 Hamburg, Germany}

\date{\today}
\begin{abstract}
Moiré heterostructures hold the promise to provide 
platforms to tailor strongly correlated and topological states of matter. Here, we theoretically propose the emergence of an effective, rectangular moiré lattice in twisted bilayers of SnS with nonsymmorphic symmetry. Based on first-principles calculations, we demonstrate that strong intrinsic spin-orbit interactions render this tunable platform a moiré semimetal that hosts 2D hourglass fermions protected by time-reversal symmetry $\mathcal{T}$ and the nonsymmorphic screw rotation symmetry $\widetilde{\mathcal{C}}_{2y}$. We show that topological Fermi arcs connecting pairs of Weyl nodal points in the hourglass dispersion are preserved for weak electron-electron interactions, particularly in regions of superconducting order that emerge in the phase diagram of interaction strength and filling. Our work established moir\'e engineering 
as an inroad into the realm of correlated topological semimetals and may motivate further topology related researches in moiré heterostructures.
\end{abstract}

\maketitle

\paragraph{Introduction ---}
Moir\'e superlattices that emerge upon twisting few layers of van der Waals heterostructures have elevated twisted graphene and transition metal dichalcogenide (TMD) multilayers to the forefront of condensed matter physics. Twistronics exploits quantum interference effects to quench kinetic energy scales of electrons moving in the effective long-ranged moiré potential~\cite{Bistritzer12233, PhysRevLett.99.256802,2009Observation,morell2010flat} such that materials can be tuned continuously to Coulomb interaction dominated regimes at low energies. This avenue allows to tailor correlated quantum phases on demand~\cite{moiresim} placing superconductors~\cite{cao2018unconventional,lu2019superconductors,yankowitz2019tuning,liu2021tuning,stepanov2020untying,arora2020superconductivity,Park2021,cao2021large,hao2021electric,kim2021spectroscopic,Chen2019ABC,park2021multi,zhang2021multi,burg2021multi,liu2022isospin,turkel2022orderly}, Mott-like correlated insulators~\cite{cao2018mott,chen2019evidence,wang2020correlated,regan2020mott}, generalized Wigner crystal states~\cite{regan2020mott,xu2020correlated,li2021imaging}, nematicity and stripe phases~\cite{Kerelsky18,Choi2019correlations,Jiang2019charge,cao2021nematicity,rubioverdu2020universal,jin2021stripe} within experimental reach.
With such advantage, twisted van der Waals materials hold the promise to be used as robust solid-state based quantum simulation platforms to realize effective correlated lattice models of different dimensionality~\cite{1dGeSe,xian2021engineering,moiresim} and offer unprecedented opportunities to study topological states in the interaction dominated regime. Recent works address the quantum anomalous Hall effects in twisted bilayer graphene and trilayer graphene/hBN superlattices driven by intrinsic strong interactions~\cite{Serlin900,nuckolls2020strongly,chen2020tunable,das2021symmetry} and the topological phase transition from a Mott insulator to a quantum anomalous Hall insulator in TMD heterostructures~\cite{li2021quantum}. Fractional Chern insulator (FCI) states are observed in twisted bilayer graphene at small magnetic field~\cite{xie2021fractional}, and even predicted to be formed at zero field in twisted graphene systems~\cite{xie2021fractional,abouelkomsan2020particle,ledwith2020fractional,repellin2020chern,liu2021gate} and TMD moir\'e superlattices~\cite{li2021spontaneous,claassen2021ultrastrong}. However, previous works mainly investigate the insulator-type of topology with strong correlations and the study of correlated topological semimetal states that can be realized in moir\'e superlattices is still missing.

In this Letter, we theoretically propose the twisted bilayer tin(II)-sulfide (tSnS) to form a rectangular moiré lattice hosting 2D hourglass fermions~\cite{2016Hourglass,Mae1602415,wang2019,jin2020} protected by the combination of time-reversal symmetry $\mathcal{T}$ and the emergent nonsymmorphic screw rotation symmetry $\widetilde{\mathcal{C}}_{2y}=\{\mathcal{C}_{2y}|0\frac{1}{2}\}$.
The two Weyl points emerging in the hourglass dispersion of strongly spin-orbit coupled tSnS are connected by Fermi arcs that counter-propagate at the system's edge and render this moiré structure a topological 2D semimetal.
The hourglass fermions are robust for a wide range of twist angles and should be accessible via gating, which enables flexible control of its electronic properties in experiments. We further demonstrate that control of their kinetic energy, i.e. the bandwidth $W$ can be achieved via twist angle variations, which allows to study the impact of electronic correlations on the topological features. We exemplify our findings using large-scale first-principles calculations from which we construct a generic and accurate tight-binding model that captures the features of the hourglass fermions in the presence of $\mathcal{T}\widetilde{\mathcal{C}}_{2y}$ and strong Rashba SOC. Treating the effect of electronic correlations within a weak-coupling random-phase approximation (RPA) approach, we study the correlated phase diagram of tSnS as a function of filling $\nu$ and interaction strength $U$. Our analysis suggests the presence of a variety of spin-/density wave (DW) and unconventional superconducting (SC) phases that originate from the specific hourglass dispersion. In particular, we show that the hourglass fermiology is robust for small interaction strengths $U \lessapprox 0.3W$ and that Fermi arc states are preserved in regions of unconventional superconducting order, which renders tSnS a topological hourglass superconductor.


\paragraph{Moiré engineering of topological 2D semimetals ---}
The idea we put forward in this work aims to expand the realm of topological 2D semimetals~\cite{jin2020,PhysRevLett.115.126803} to the field of moiré engineering. To construct a rectangular moiré superlattice, we start from chemically stable and experimentally available van der Waals compounds made of orthorhombic group-IV monochalcogenides (MXs; M=Sn, Ge; X=S, Se) which themselves feature a rectangular unit cell. 
The latter are reported to be promising nanoelectronic materials with a layered structure similar to phosphorene (space group $Pmn2_1$), which show emergent multiferroicity with a large spontaneous polarization~\cite{doi:10.1021/acs.nanolett.6b00726,2017IV}. Among the MXs, SnX exhibits higher chemical stability than GeX~\cite{doi:10.1021/acsami.6b16786} and the ferroelectric Curie temperature of SnS is higher~\cite{PhysRevLett.117.097601}. These physical properties render SnS an ideal choice for moiré engineering.
\begin{figure}
    \includegraphics[width=1\linewidth]{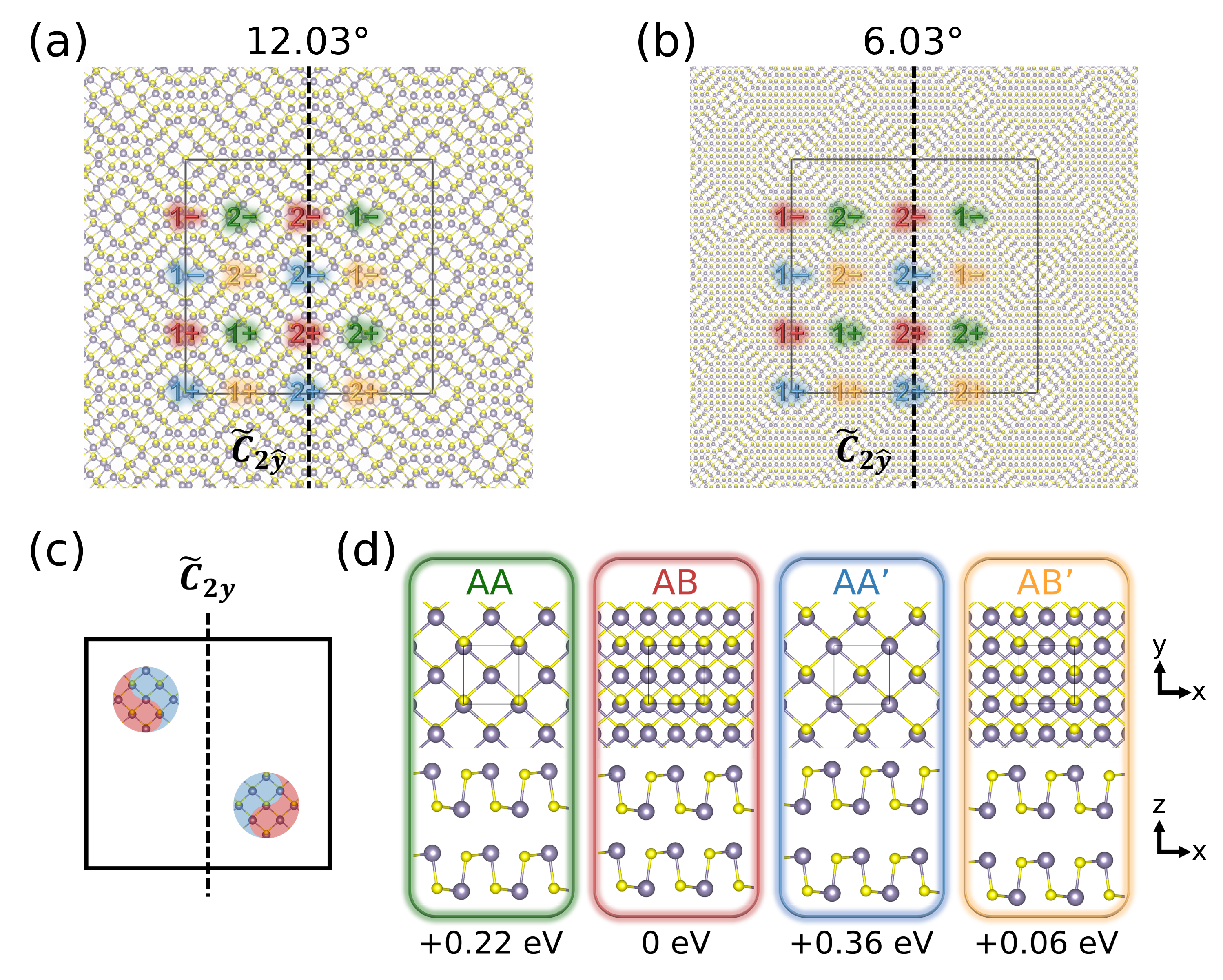}
    \caption{\label{fig1} Moiré pattern of bilayer SnS with a relative twist angle of 12.03$\degree$ (a) and 6.03$\degree$ (b). Silver and yellow spheres indicate Sn and S atoms, respectively. Stacking domains are highlighted by green (AA), red (AA’), blue (AB), and orange (AB’) numbered areas respectively, where "1/2" indicating nonequivalent stacking domains and "+/-" signs showing the symmetric domain pairs. (c) Schematic illustration of the nonsymmorphic $\widetilde{\mathcal{C}}_{2y}$ connects two AA domains. The translucent Taiji symbols are plotted to display the rotational symmetry. (d) Top and side views of the four stacking structures of bilayer SnS. The unit cell is marked by black solid rectangles in the upper panel.}
\end{figure}
We first test this idea by performing \textit{ab initio} calculations of two layers of SnS with a twist angle of $\theta = 12.03\degree$. This is the smallest cell with negligible strain ($<0.05\%$ on $y$ axis) to set up a commensurate structure. The atomic structure is relaxed to minimize the energy and stress by density functional theory (DFT), and the optimized structure retains a rectangular lattice as demonstrated in Fig.\ref{fig1}~(a). The moiré pattern exhibits a continuous transformation among four stacking domains shown in Fig.\ref{fig1}~(d). Since AA' and AB' stacking domains in parallel bilayer SnS are energetically favored, they grow larger with decreasing twist angle, as displayed in Fig.~\ref{fig1}~(b) for $\theta=6.03\degree$. The superlattice belongs to space group $P2_1$ (No.4) with only one nonsymmorphic symmetry $\widetilde{\mathcal{C}}_{2y}$ consisting of a twofold rotation $\mathcal{C}_{2y}$ in conjunction with translation by half a Bravais lattice vector along the axis of rotation, as illustrated in Fig.\ref{fig1}~(c).  A closer view of symmetry operations in 6.03 \degree structure can also be found in the Supplemental Material (SM). Disregarding spin-orbit coupling (SOC) in our calculations first, the twisted system has a band gap of 1.18 eV and develops flat bands near the edges of valence and conduction bands, as shown in Fig.~\ref{fig2}~(a). The bands near the valence band edge show four-fold Dirac nodal lines along $Y-S$ and are separated into two sets which are marked in Fig.~\ref{fig2}~(a). Charge densities for set I and II are mainly localized on two inequivalent AA stacking domains [see Fig.~S3~(a) in SM] with negligible residual coupling. Therefore, these stacking domains can be regarded as effective localized superatoms forming the low-energy bands in tSnS. In fact, two AA domains in each set of bands are connected by the two-fold screw axis $\widetilde{\mathcal{C}}_{2y}$ that hence protects the Dirac nodal line in the band structure. When artificially shifting the top layer of tSnS along $y$ to break $\widetilde{\mathcal{C}}_{2y}$, the Dirac nodal lines for both band sets are destroyed, as displayed in Fig.~\ref{fig2}~(c). Furthermore, we show that these bands are robust against twist angle variations. By decreasing the twist angle, similarly-shaped bands with reduced bandwidth $W$ appear near the valence band edge, as shown for $\theta=6.03\degree$ in Fig.~\ref{fig2}~(b).

\begin{figure}
    \includegraphics[width=1\linewidth]{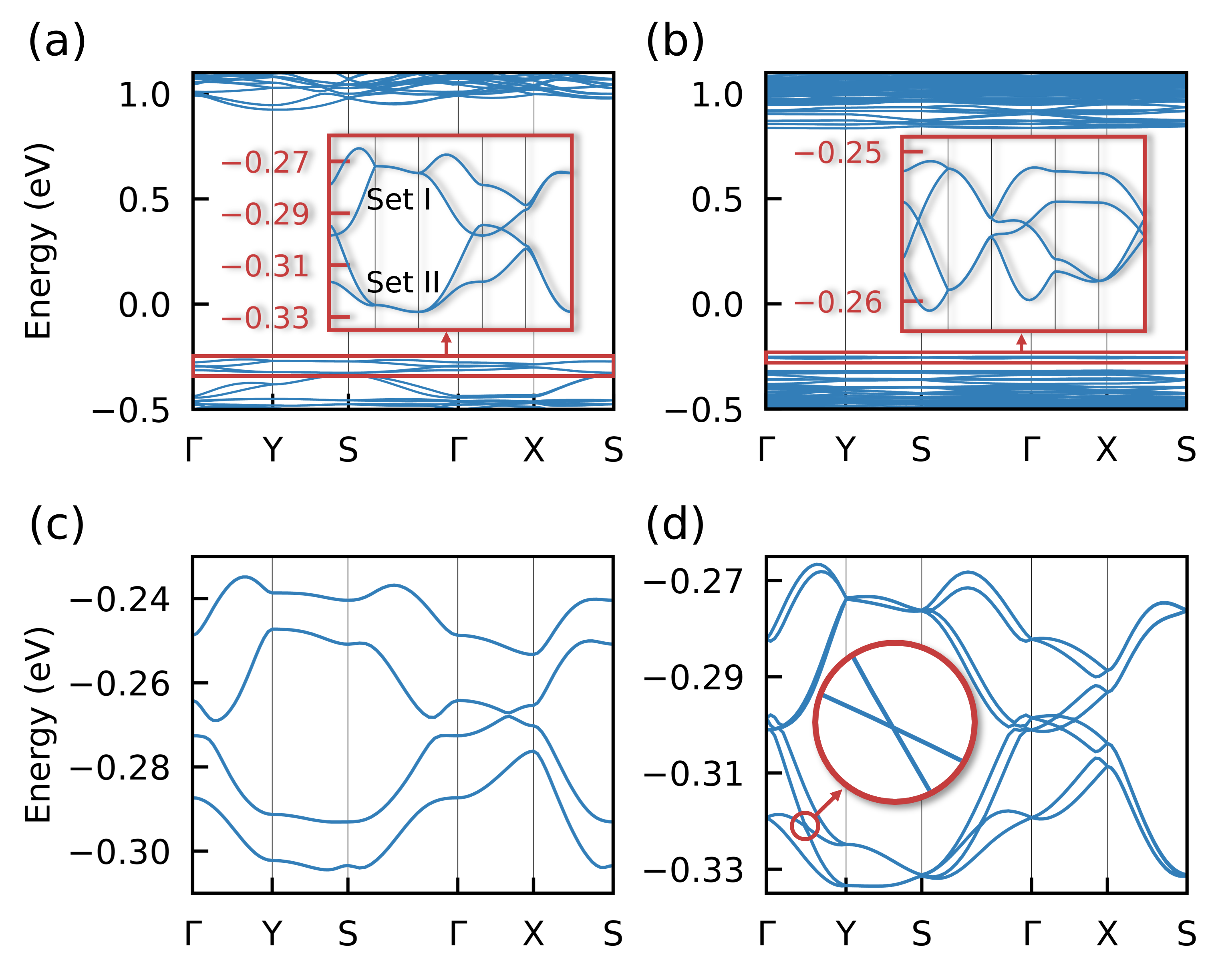}
    \caption{\label{fig2} (a) (b) Band structures calculated by DFT without SOC for twisted bilayer SnS with a relative twist of 12.03$\degree$ and 6.03 $\degree$, respectively. The inset figures show the bands (defining bandsets I and II) at the valence edge highlighted as red boxes. (c) The valence bands (without SOC) of a top layer shifted structure without $\widetilde{\mathcal{C}}_{2y}$ with a relative twist of 12.03$\degree$. (d) DFT calculated valence bands for the 12.03$\degree$ structure with SOC included. The inset highlights the hourglass-like crossing for bandset II in $\Gamma$-$Y$.
    }
\end{figure}

\begin{figure}
    \includegraphics[width=1\linewidth]{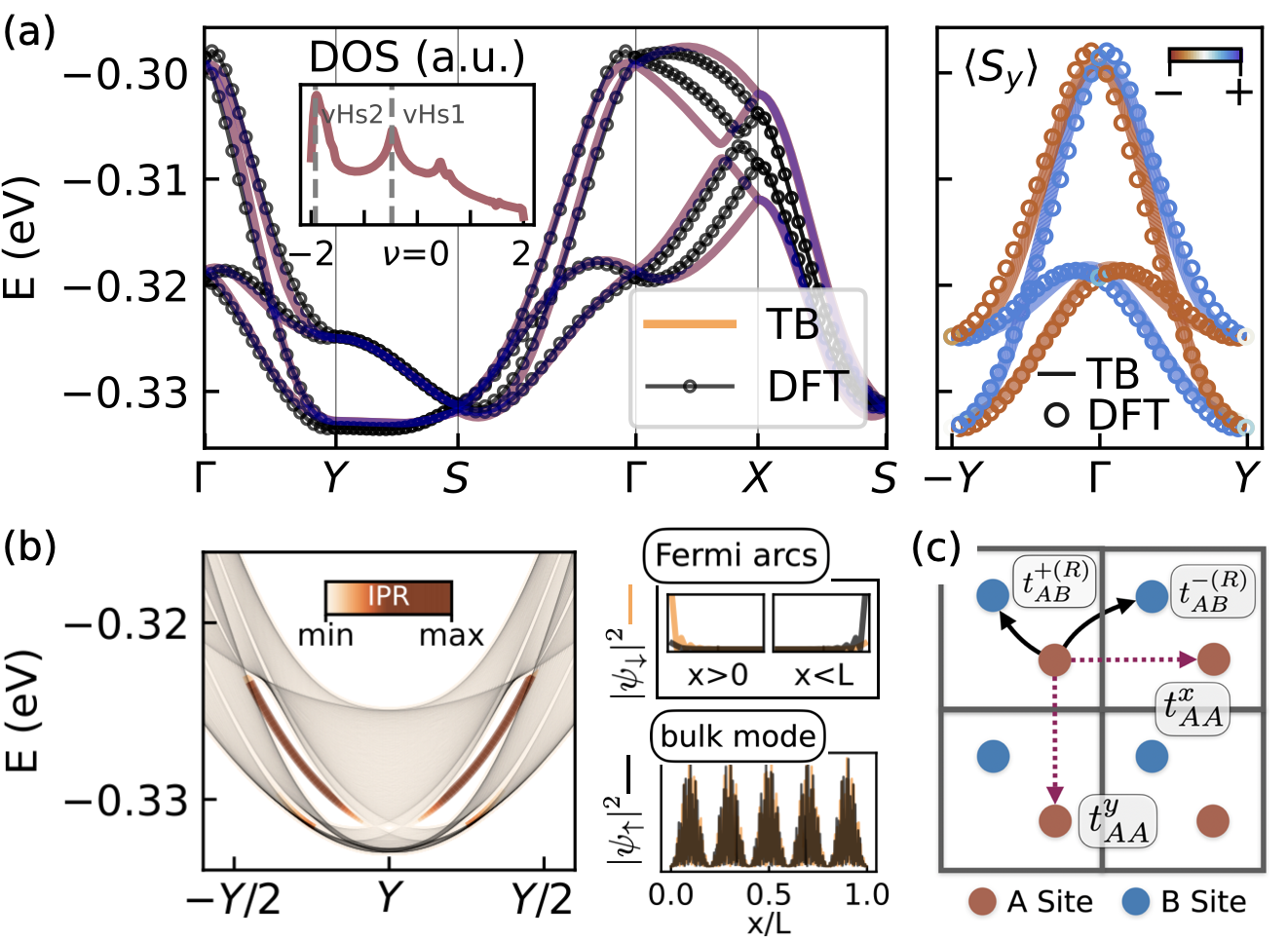}
    \caption{(a) Hourglass-like band structure and spin-momentum locking in the low-energy bands of twisted SnS. Orange and black lines in the left panel indicate the dispersion obtained from DFT and the TB fit, respectively. The inset shows the density of states including van-Hove singularities vHs1/vHs2 that arise from saddle points of the dispersion at $Y$. The combination of strong intrinsic spin-orbit coupling and $\mathcal{T}\widetilde{\mathcal{C}}_{2y}$ gives rise to a pair of Weyl nodal points at $\bvec k = \pm Y/2$ with opposite spin-$y$ polarization. (b) In a slab geometry of tSnS with open boundaries in $x$-direction, the pair of Weyl points is connected by two counter-propagating Fermi arcs that are localized at opposite boundaries as demonstrated by the high inverse participation ratio (IPR) and the wave function profile. (c) Sketch of the effective TB model to mimic the dispersion of the rectangular moiré lattice. To restrict the symmetry of the system to $\widetilde{\mathcal{C}}_{2y}$, we account for spatial anisotropy $(+/-)$ and include up to 5th nearest-neighbor kinetic hopping terms, see SM. $SU(2)$ symmetry is broken in the presence of strong Rashba SOC as indicated by $t_{AB}^{+R}$.}
\label{fig3} 
\end{figure}

Including spin-orbit interactions leads to significantly altered band degeneracies, as shown in Fig.~\ref{fig2}~(d). Time reversal symmetry enforces Kramers pairs at $\mathcal{T}$-invariant momentum points and the four-fold degenerate Dirac nodal line along $Y-S$ splits into two-fold degenerate Weyl nodal lines. In particular, we observe hourglass-shaped bands along $\Gamma-Y$ in bandset II characterized by the formation of a Weyl nodal point that is protected in the presence of SOC, as shown in the inset of Fig.~\ref{fig2}~(d). Similar hourglass features are observed along $X-S$ in bandset I/II, albeit the band splitting at the $S$ point of $\sim$0.2 meV is considerably smaller, see SM for more details. We note that the band crossings along $\Gamma-S$ are accidental as they are not protected by any symmetry. Our first-principles calculations further suggest rigid spin-momentum locking in the low-energy bands of tSnS. The nonsymmorphic symmetry $\widetilde{\mathcal{C}}_{2y}$ mandates that the electron's spin along $\Gamma-Y$ is solely polarized in the $y$-direction, as shown in the 
right panel of Fig.~\ref{fig3}~(a). Similar spin textures have been theoretically studied in phosphorene~\cite{PhysRevB.94.205426} and Bi/Cl-SiC(111)~\cite{wang2019}.

\paragraph{Effective tight-binding Hamiltonian and hourglass fermiology ---}
The emergence of charge localization points in tSnS hints towards an effective description of the low-energy valence band manifold within a tight-binding (TB) approach. Following the real-space distribution of the wave functions obtained from DFT, we adapt a rectangular lattice model with two sublattices $A(B)$, as shown in Fig.~\ref{fig3}~(c). In the presence of spin-orbit coupling, the symmetry operators of this model are $\mathcal{T} = -i \sigma_y \mathcal{K}$ and $\widetilde{\mathcal{C}}_{2y}= i \tau_x \sigma_y$, in terms of Pauli matrices $\sigma$ ($\tau$) acting on the spin (sublattice) degree of freedom and $\mathcal{K}$ is the operator of complex-conjugation. To conserve the non-symmorphic symmetry $\widetilde{\mathcal{C}}_{2y}$ and to account for the anisotropy in $x$ and $y$ that manifests due to the presence of the effective rectangular moiré unit cell, we allow for anisotropic hopping amplitudes to the right/left from each reference site as indicated by $+/-$ in Fig.~\ref{fig3}~(c). 

The strong intrinsic spin-orbit coupling of tSnS is captured by Kane-Mele and Rashba SOC terms. The latter particularly break inversion and $SU(2)$ symmetry completely such that the non-interacting Hamiltonian reads
\begin{equation}
H^0 = \sum_{ij s} t_{ij}^{\text{kin}} e^{i \phi_{ij} s} c^{\dagger}_{i s} c^{\phantom \dagger}_{js} + i \sum_{ij s s^{\prime}} t_{ij}^{R} c^{\dagger}_{i s} \left ( \bvec\sigma \times \bvec d_{ij} \right )_z c^{\phantom \dagger}_{j s^{\prime}},
\label{eq:tb_general}
\end{equation}
where $c^{(\dagger)}_{is}$ creates (annihilates) an electron with spin $s$ on moiré site $\bvec{r}_i$ that belongs to one of the sublattices $o \in \{A,B \}$. To achieve sufficient agreement with the \textit{ab intio} DFT band structure and mimic the correct spin texture, we allow for nearest-neighbor SOC terms and up to $5^{\text{th}}$ nearest-neighbor kinetic hopping terms\footnote{A simple Python code for the TB model is available online: \url{https://git.rwth-aachen.de/ammon.fischer/tight-binding-model-twisted-sns}}, see SM for details. The success of fitting the flat bands in tSnS within our 13-parameter tight-binding model is demonstrated in panel (a). The excellent agreement between TB and first-principles calculations underlines the presence of an effective nonsymmorphic symmetry $\widetilde{\mathcal{C}}_{2y}$ leading to the topological realization of tunable hourglass fermions via moiré engineering in tSnS.

The origin and robustness of the nodal lines and Weyl points in the band structure can be understood from a symmetry perspective~\cite{PhysRevLett.115.126803}. The nonsymmorphic symmetry $\widetilde{\mathcal{C}}_{2y} = \{\mathcal{C}_{2y}|0\frac{1}{2}\}$ protects additional degeneracies along invariant lines in momentum space that satisfy $\mathcal{C}_{2y} \bvec k = \bvec k$, i.e. along $\Gamma/X-Y/S$. In this invariant space, Bloch states can be chosen as eigenstates of $\widetilde{\mathcal{C}}_{2y} |u_{\bvec k} \rangle = \pm \lambda e^{ik_y/2}|u_{\bvec k} \rangle$. In the absence of spin-orbit coupling, i.e. [$\mathcal{T}^2=1$; $(\mathcal{C}_{2y})^2=1$; $\lambda=\pm 1$], the product of time-reversal and non-symmorphic symmetry fulfills $(\mathcal{T}  \widetilde{\mathcal{C}}_{2y})^2=-1$ at $k_y=\pi$, resulting in a Kramers degenerate Dirac nodal line along $Y-S$. Upon including SOC, i.e. [$\mathcal{T}^2=-1$;  $(\mathcal{C}_{2y})^2=-1$; $\lambda=\pm i$], the two double degenerated Weyl nodal lines along $Y-S$ are still protected by $(\mathcal{T}\widetilde{\mathcal{C}}_{2y})^2=-1$ at $k_y=\pi$. At the time reversal invariant momentum points $\Gamma/X$ ($Y/S$), $\widetilde{\mathcal{C}}_{2y}$ has eigenvalues $\pm i$ ($\pm 1$). Since $\mathcal{T}$ is anti-unitary and commutes with $\widetilde{\mathcal{C}}_{2y}$, the Kramers partner has eigenvalues $\widetilde{\mathcal{C}}_{2y} \left( \mathcal{T} |u_{\bvec k} \rangle \right ) = \mp i e^{-ik_y/2}\left ( \mathcal{T} |u_{\bvec k} \rangle \right )$. Therefore, Kramers partners have opposite eigenvalues $\pm i$ at $\Gamma/X$, but the same eigenvalues $\pm 1$ at $Y/S$ resulting in the formation of a Weyl point and the hourglass-like dispersion along $\Gamma-Y$ and $X-S$.

The topological character of the Weyl points manifests in the existence of two Fermi arcs when extending the system to a semi-infinite ribbon with open boundary conditions in $x$ direction, as shown in Fig.~\ref{fig3}~(b). The Fermi arcs connect the pair of Weyl points along $Y-\Gamma-Y$, have opposite spin and counter-propagate along the opposite boundary indicated by the high inverse participation ratio $\text{IPR}(\bvec k, b) = \sum_{o,s} |u_{os,b}(\bvec k)|^4$ and the wavefunction profile.

\paragraph{Phase diagram of the moir\'{e} Hubbard model of tSnS ---}
The advantage of realizing topological 2D semimetals via moiré engineering lies in the control over the kinetic energy scales $W(\theta)$ as function of the twist angle, which may eventually drive the system into interaction dominated regimes at low energies. To investigate the effect of correlations and characterize putative low-temperature phases and their impact on the Fermi arcs, we study a moir\'{e} Hubbard model~\cite{andy2021hartree,kleblWSe2} by complementing the tight-binding model Eq.~\eqref{eq:tb_general} with a repulsive on-site term
\begin{equation}
    H^U = U\,\sum_{i} n_{i,\uparrow} n_{i,\downarrow} \,.
\label{eq:moire_U}
\end{equation}
This particular choice of the interaction does not capture the entire long-ranged nature of the Coulomb interaction. Nevertheless, long-ranged contributions to the interaction may be screened in a controlled manner by substrate engineering~\cite{stepanov2020untying,liu2022isospin} and previous works addressing twisted TMDs with strong spin-orbit coupling, e.g. tWSe$_2$~\cite{andy2021hartree,wu2018hubbard}, indicate that important interaction effects stem from on-site repulsion.

\begin{figure}
    \centering
    \includegraphics[width=\columnwidth]{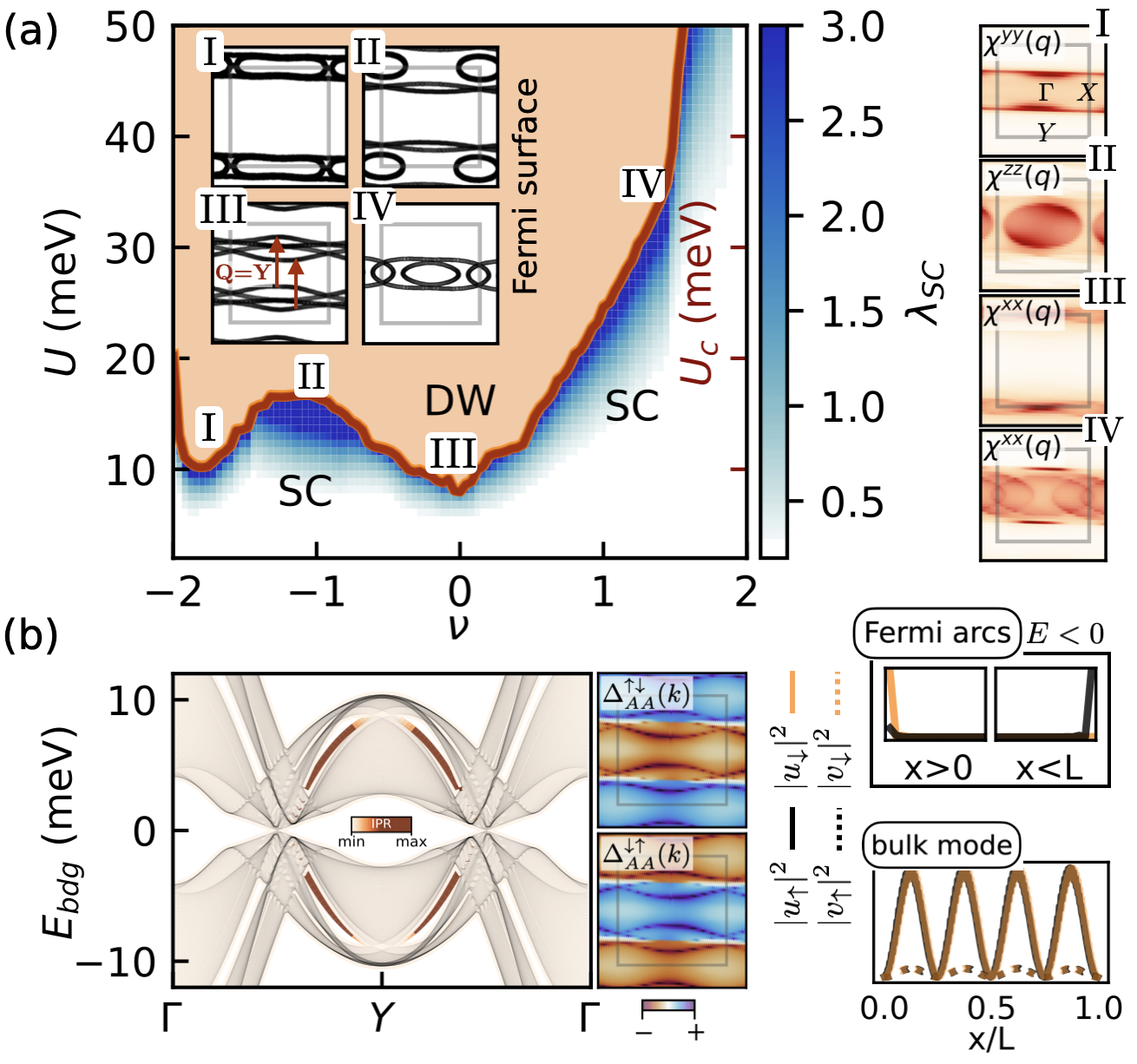}
    \caption{Phase diagram obtained from the multi-orbital RPA analysis of tSnS. Mixed spin-/density waves (DW, orange) and superconducting instabilities (SC, blue) are shown as function of the moir\'{e} Hubbard-$U$ and filling $\nu$. Different DW/SC regions are separated by the critical scale $U_c$ (red line) that marks the onset of DW order. At the filling of the Weyl nodal point (III), commensurate DW instabilities with ordering vector $\bvec q = Y$ (stripe-$y$ order) emerge at interaction scales $U_c \approx 0.3 \,  W$ relative to the bandwidth $W$, driven by the vicinity to vHs1 and by nesting features of the Fermi surface as shown in the inset of (a), $I-IV$. The corresponding spin-/charge response has maximal weight in the  $\chi^{xx}(\bvec q)$ channel as shown in the left panel. (b) Fluctuations around the stripe-$y$ ordered state give rise to mixed-parity superconducting instabilities $\Delta_{g(u)}$ that transform even/odd under the nonsymmorphic symmetry $\widetilde{\mathcal{C}}_{2y}$. The leading SC state with highest transition temperature $T_c \propto \lambda_{SC}$ for all fillings $\nu$ is $\Delta_{g}$, which is nodal and preserves the Fermi arcs at $k = Y/2$. In the Bogoliubov de-Gennes (BdG) quasi-particle band structure of a slab geometry with open boundaries in $x$ direction, these modes are doubled due to emergent particle-hole symmetry and are highly localized at the edge of the system as shown by the particle- and hole wave functions $|u|^2$, $|v|^2$ in the right panel.}
    \label{fig4}
\end{figure}

To analyze electronic instabilities within the weak-coupling limit, we resort to the multi-orbital random-phase approximation (RPA)~\cite{graser2009near,fischer2021unconventional,romer2021superconducting,fischer2021spin}. By expanding the two-particle Green's function $\chi^{\text{RPA}}(\bvec q)$ in a series of bubble- and ladder diagrams to infinite order in $U$ and rewriting the renormalized interactions in the particle-hole (ph) and particle-particle (pp) channel, tendencies towards either spin-/charge density wave order (DW) or superconducting order (SC) are resolved, see SM. Technically, these phases are separated by the critical scale $U_c$ that marks divergences in the ph-channel and thus indicates an instability towards DW order. As spin-/charge density instabilities generally mix in the absence of $SU(2)$ symmetry, we analyze the physical susceptiblities~\cite{scherer2018spin-orbit,kleblWSe2} given by $\chi^{ij}(\bvec q) = \sigma_{i} \chi^{\text{RPA}} \sigma_{j}$, where $i,j \in \{0,x,y,z\}$ denote the charge- and spin correlations in the respective lattice direction. For interaction values $U<U_c$, the effective coupling in the pp-channel $\Gamma^{\text{pp}}(\bvec k, \bvec k^{\prime})$ further allows to track pairing instabilities by solving the linearized gap equation $\lambda_{SC} \hat\Delta(\bvec k) = \hat{\Gamma}^{pp}(\bvec k, \bvec k^{\prime}) \circ \hat\Delta(\bvec k^{\prime})$. The largest eigenvalue $\lambda_{SC}$ corresponds to the highest critical temperature $T_c$ and the corresponding eigenfunction $\hat\Delta(\bvec k)$ yields the symmetry of the dominating SC order parameter. 

We outline correlated phases emerging from electronic interactions in tSnS in a phase diagram of interaction strength $U$ and filling $\nu$ as shown in Fig.~\ref{fig4}~(a). The particular $\Lambda$-shape of the hourglass dispersion along $Y-\Gamma-Y$, see Fig.~\ref{fig3}~(a), causes nesting features of the Fermi surface with connecting vector $\bvec Q \propto Y$ for wide filling regions as well as two van-Hove singularities vHs1/vHs2 that emerge at saddle points of the dispersion at $Y$. This facilitates the formation of mixed charge-/spin density waves (DW) with momentum transfer $\bvec q = \bvec Q$, whereas the physical response channel $\chi^{ij}(\bvec q)$ is mandated by the spin polarization of the respective Fermi surface sheets, see SM. The critical scale $U_c$ reaches a minimum at vHs2 ($\nu \approx -2$), where the Fermi surface consists of pockets along $Y-S$ with strong $S_y$ polarization. This enhances the spin-spin response in the $\chi^{yy}$ channel with ordering vector $\bvec q = Y/2$ as shown in the right panel of Fig.~\ref{fig4}, I. Close to half-filling and vHs1, the Fermi surface is additionally nested as it consists of two broadened lines that are connected by the vector $\bvec Q = Y$, see inset of Fig.~\ref{fig4}, III. Hence, dominant weight in the response matrix is found in the $\chi^{xx}(\bvec q)$ channel with peaks at $\bvec q = Y$ indicating commensurate stripe-$y$ order. On the electron-doped side, the critical scale $U_c$ raises as the bands along $\Gamma-Y$ become steeper and hence diminish the electronic ordering tendencies of the system. As any kind of mixed spin-/charge wave order will inevitably break time-reversal symmetry $\mathcal{T}$, the hourglass dispersion will be gapped at the interaction scale $U_c$. At the filling of the Weyl nodal point, the critical scale takes values of $U_c \approx W/3$, where $W=30\,\text{meV}$ denotes the bandwidth of the hourglass fermion. 

When doping away from the vHs, DW order is steadily replaced by mixed-parity SC instabilities on the electron- and hole doped side, see Fig.~\ref{fig4}~(a). The SC order originates from fluctuations around the stripe-$y$ order and can be classified according to whether it transforms even/odd under $\widetilde{C}_{2y}$, i.e. $U^{\dagger}(\bvec k) \Delta (\bvec k) U(\bvec k) = \pm \Delta (\bvec k)$ with the unitary matrix $U(\bvec k) = i \tau_x \sigma_y e^{ik_y/2} \mathcal{R}_{k_x \to -k_x}$. The leading (sub-leading) solutions are labeled as $\Delta_{g(u)}$ respectively and their momentum dependence in the intra-sublattice, opposite-spin component $\Delta_{AA}^{\uparrow\downarrow}(\bvec k)$ resembles a $d_y \, (p_x)$-wave with nodal lines in $x \, (y)$ direction as shown in Fig.~\ref{fig4} for $\Delta_{g}$. For information about the subleading instability, the reader may refer to the SM. In particular, the SC nodal lines of both possible instabilities happen to coincide with the position of the Weyl nodal point at $\bvec k = Y/2$ such that the hourglass topology is preserved in the presence of SC order up to the critical scale $U_c$. Hence, we argue that all unconventional SC instabilities arising from repulsive interactions in tSnS preserve the Fermi arc states that connect the pair of Weyl points and 
thus preserve the non-trivial topology of the non-interacting electronic band structure. 
In the slab geometry with open boundary conditions in $x$ direction, the Bogoliubov de-Gennes (BdG) quasiparticle bands indicate the intact Weyl point featuring localized Fermi arcs that are doubled in terms of particle- and hole degrees of freedom $|u_{\uparrow(\downarrow)}|^2$, $|v_{\uparrow(\downarrow)}|^2$ due to emergent ph-symmetry in the SC phase. Indeed, the wavefunctions profiles in Fig.~\ref{fig4} clearly show edge localization, whereas ordinary bulk modes are distributed over the whole length of the slab.
The sub-leading instability $\Delta_u$, see SM, further features a zero-energy Majorana mode whose particle-/hole amplitudes are localized at opposite boundaries. 

\paragraph{Discussion} --- 
In this work, we predict the robust realization of an effective rectangular moiré lattice model with strong Rashba spin-orbit coupling via moiré engineering of tSnS. Based on our large scale first-principles calculations, the limited crystalline symmetries of the twisted compound restrict the symmetry group of the effective model to the nonsymmorphic screw rotation symmetry $\widetilde{\mathcal{C}}_{2y}$ and time-reversal symmetry $\mathcal{T}$. This leads to symmetry protected pairs of Weyl nodal points along $\Gamma-Y$ that are connected by Fermi arcs, which renders tSnS a 2D topological moiré semimetal. The hourglass fermions are robust against twist angle variations and in particular in the presence of weak interactions. While mixed charge-/density waves (DW) break $\mathcal{T}$-symmetry and hence potentially destroy the Weyl nodal points, our analysis suggests that Fermi arc states are preserved in regions of the electronic phase diagram governed by SC instabilities. We hence argue that when starting from purely repulsive on-site interactions, any kind of unconventional SC order mediated by spin-/charge fluctuations inevitably will be of topological nature and will preserve the non-trivial topology of the hourglass fermions. Hence, our approach opens a rich playground towards the experimental realization of tunable Weyl-like physics, particularly as SnS is chemically stable and the valence band manifold should be accessible via gating. An interesting observation of our weak-coupling analysis concerns the observation of Majorana zero energy modes in sub-leading order. This hints towards a meta-stable state of the system that could potentially be stabilized by e.g. out-of-equilibrium dynamics~\cite{claassen2019universal}, light-driving in a cavity \cite{doi:10.1063/5.0083825} or by applying strain or an electrical field to the sample. 
Our work hence elevates twisted systems to include the broad realm of topological semimetals and may motivate further topology and valleytronics related researches in moiré heterostructures.

\begin{acknowledgments}
L.X. acknowledges the support by the National Key R \& D Program of China (2021YFA1202902) and the Key-Area Research and Development Program of Guangdong Province of China (Grant No. 2020B0101340001). A.F, L.K and D.M.K acknowledge funding by the Deutsche Forschungsgemeinschaft (DFG, German Research Foundation) under RTG 1995, within the Priority Program SPP 2244 ``2DMP'' and under Germany's Excellence Strategy - Cluster of Excellence Matter and Light for Quantum Computing (ML4Q) EXC 2004/1 - 390534769. We acknowledge computational resources provided by RWTH Aachen University under project number rwth0763. This work was supported by the Max Planck-New York City Center for Nonequilibrium Quantum Phenomena. Work at SUSTech was supported by the National Natural Science Foundation of China under Grant No. 11774142, and the Shenzhen Basic Research Fundunder Grant. No. JCYJ20180504165817769. The computer time was supported by the Center for Computational Science and Engineering of Southern University of Science and Technology. This work was supported by the European Research Council (ERC-2015-AdG694097), the Cluster of Excellence ‘Advanced Imaging of Matter' (AIM), Grupos Consolidados (IT1249-19) and Deutsche Forschungsgemeinschaft (DFG) -SFB-925–project 170620586.
\end{acknowledgments}

\bibliography{References.bib}
\end{document}


\title{Supplementary Material for:\\Moir\'e Engineering of Nonsymmorphic Symmetries and Hourglass Superconductors}
\date{\today}
\maketitle
\tableofcontents
\newpage
 
\section{Computational methods}
The first-principles calculations are performed within the density functional theory (DFT) framework using the projected augmented-wave (PAW) method\cite{PhysRevB.50.17953} as implemented in the Vienna ab initio simulation package (VASP)\cite{KRESSE199615,PhysRevB.54.11169}. The exchange-correlation term is treated in the generalized gradient approximation (GGA) of Perdew-Burke-Ernzerhof (PBE)\cite{PhysRevLett.77.3865}. The Kohn-Sham orbitals are expanded in a plane wave basis set with an energy cutoff of 350 eV. A 1$\times$1$\times$1 k-point grid is used for the ground state and relaxation calculations. The experimental lattice constants for bulk SnS ($a$=4.07 {\AA}, $b$=4.31{\AA}) are used to build the supercell of twisted bilayer SnS. A negligible strain ($\approx$ 0.05\%) is applied on $b$ to satisfy the commensurate periodic boundary condition. All the structures are optimized until the Hellman-Feynman forces on each atom are smaller than 0.01 eV/{\AA} and the energy convergence is reached when the energy difference is below 10$\rm{^{-6}}$ eV between two consecutive self-consistent steps. A vacuum space of more than 15 {\AA} along the $z$ direction is built to decouple artificial interactions between periodic neighboring slabs. Van der Waals corrections are taken into account using the DFT-D2 method of Grimme\cite{doi:10.1063/1.3382344}. The VESTA\cite{Momma:db5098} package is used to visualize the geometry structures and the real space illustration of wave functions. The twisted bilayer SnS supercells with twist angles of 12.03\degree and 6.03\degree contain 728 and 2888 atoms, respectively.
\newpage

\section{Illustration of stacking domains under the nonsymmorphic symmetry}
\begin{figure}[ht]
    \includegraphics[width=0.8\linewidth]{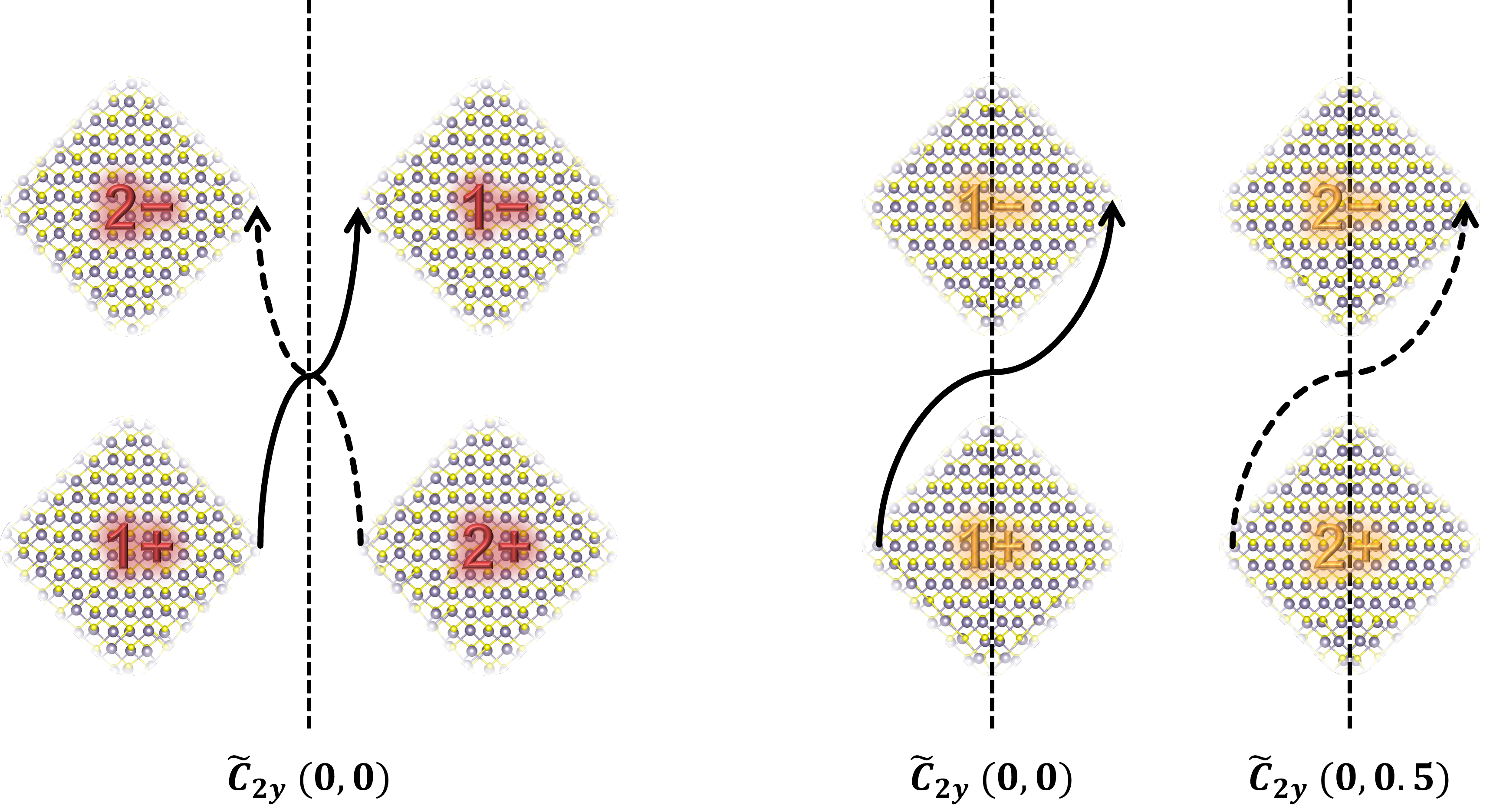}
    \caption{Schematics diagrams of the screw rotation symmetry connecting two AB (left panel) and AB' (right panel) stacking domains for tSnS with a relative twist of 6.03$\degree$. The straight dashed lines represent screw axes. The solid and dashed curved arrow lines show the symmetry operation in 2 inequalent domain pairs within the unit cell. \label{fig:symmetry}}
\end{figure}

The unitcell of a tSnS structure contains 4 different domains for each type of stacking. The screw rotation $\widetilde{\mathcal{C}}_{2y}$ further divides them into two pairs, as displayed in Fig.~\ref{fig:symmetry}. The stacking domains labeled by "1" and "2" are totally different, while "+" and "-" indicate screw mapping. The atoms in the top layer of "+" are mapped into the lower layer "-" domains. According to the position of their domain centers relative to the screw axes, the arrangement these domains along $y$ can be either zigzag or linearly alternate, as shown in the left and right panel of Fig.~\ref{fig:symmetry} respectively.

\newpage

\section{DFT calculated band structures}
\begin{figure}[ht]
    \includegraphics[width=0.8\linewidth]{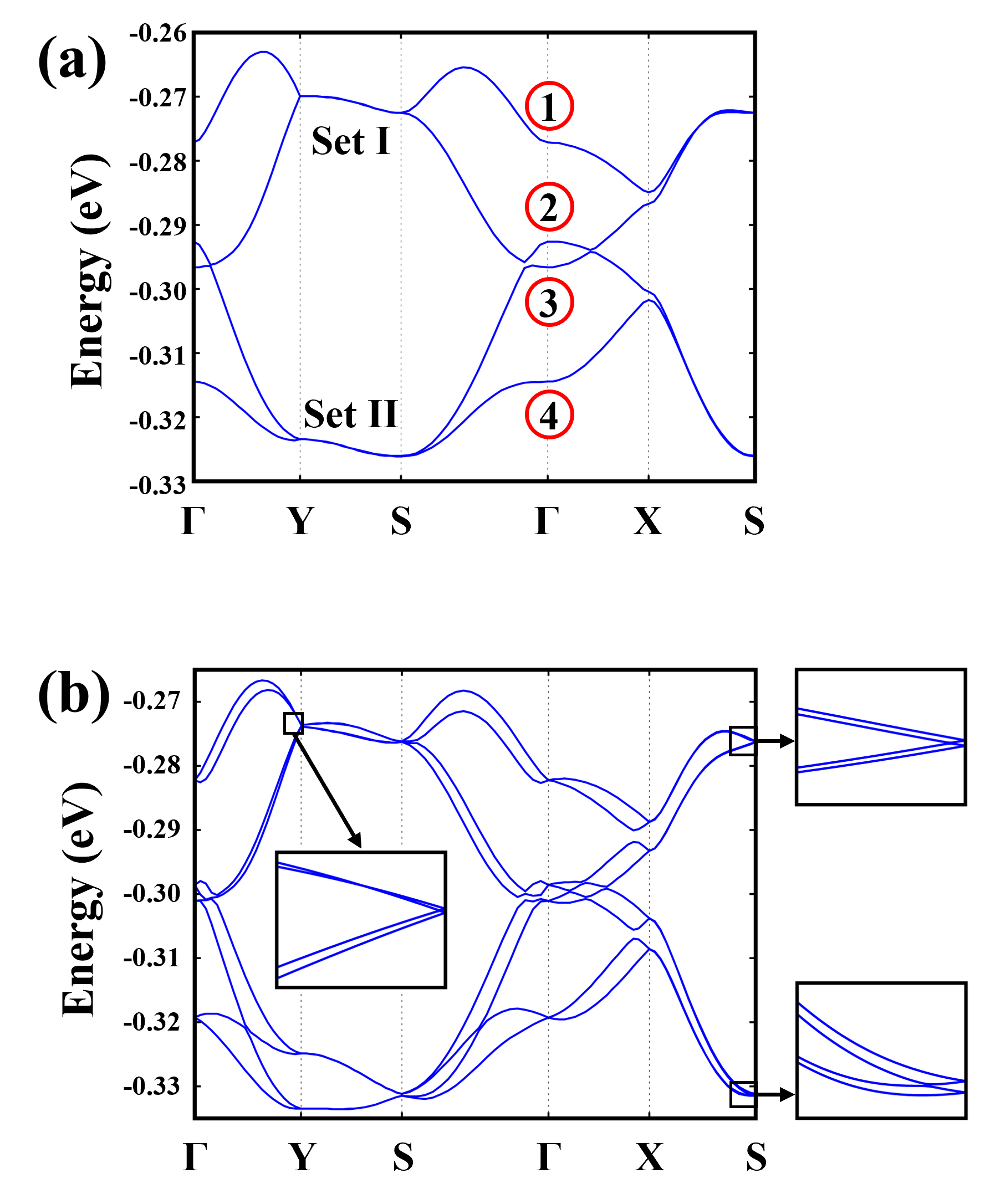}
    \caption{ Band structures near VBM calculated by DFT (a) with and (b) without SOC for tSnS with a relative twist of 12.03$\degree$. The insets in (b) show the symmetry protected band crossings near $Y$ and $S$.\label{fig:dftband}}
\end{figure}
\newpage

\section{Real space illustration of wave functions}

\begin{figure}[ht]
    \includegraphics[width=0.8\linewidth]{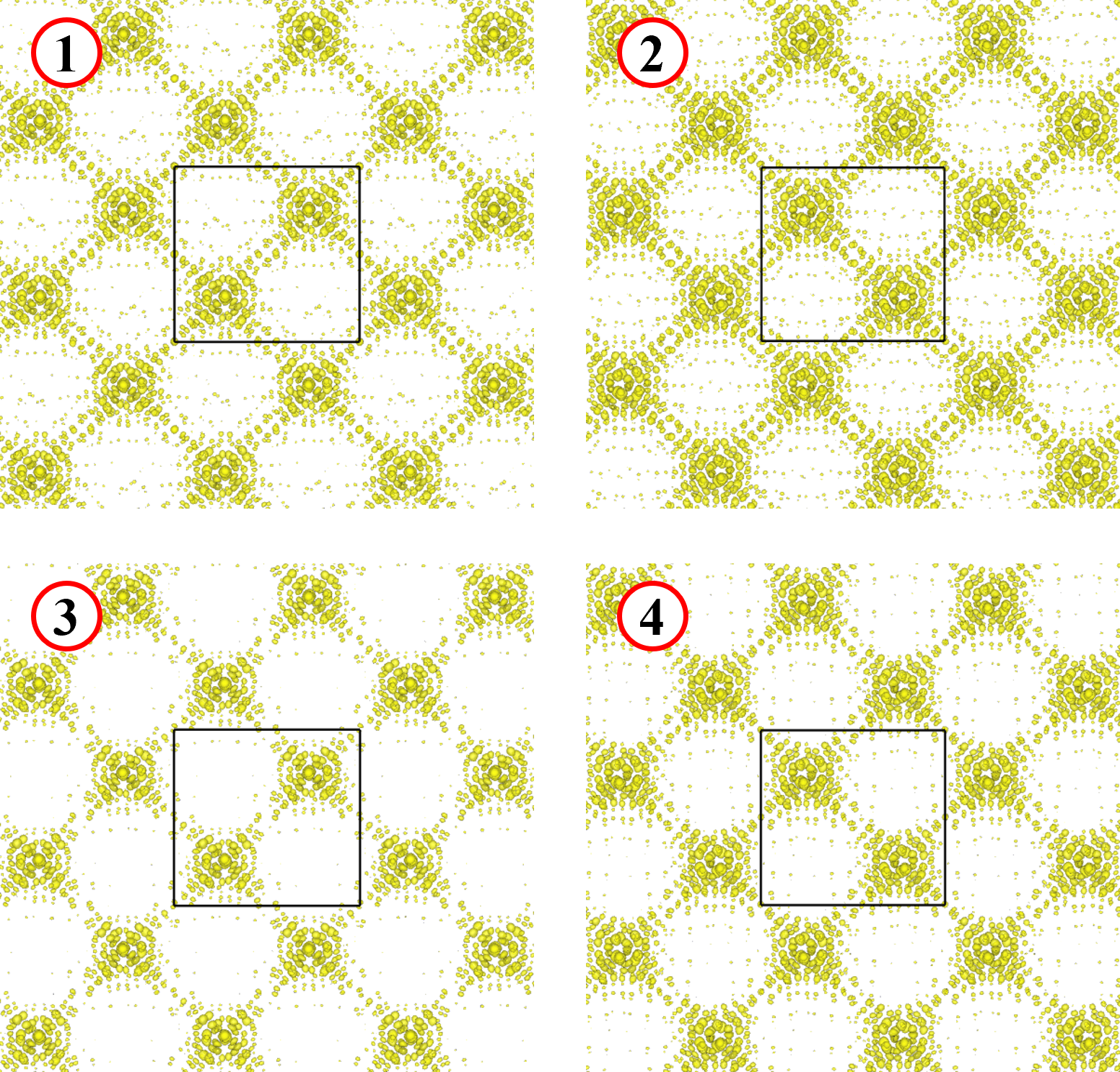}
    \caption{ Real space illustration of wave function squared at $\Gamma$ point for bands labeled by 1-4 in Fig.~\ref{fig:dftband} (a).
    \label{fig:wavefunction}}
\end{figure}
\newpage

\section{Low-energy tight-binding model}
The emergence of strong charge localization points as shown in Fig.~\ref{fig:wavefunction} hints towards an effective description of the low-energy valence band manifold (VBM) within a tight-binding (TB) approach. Following the real-space distribution of the wave functions obtained from first-principle calculations, we adapt a rectangular lattice model with two sites per unit cell, in the following referred to as sublattice A and B with relative coordinates $\bvec r_A = (\frac{3}{4}, \frac{1}{4})$ and $\bvec r_B = (\frac{1}{4}, \frac{3}{4})$. The wavefunctions obtained from DFT calculations suggest that the residual coupling between the upper (lower) 4 bands in the VBM is rather small and we therefore model each of these bands separately resulting in the two band sets I/II. 

\begin{figure*}
    \includegraphics[width=1\linewidth]{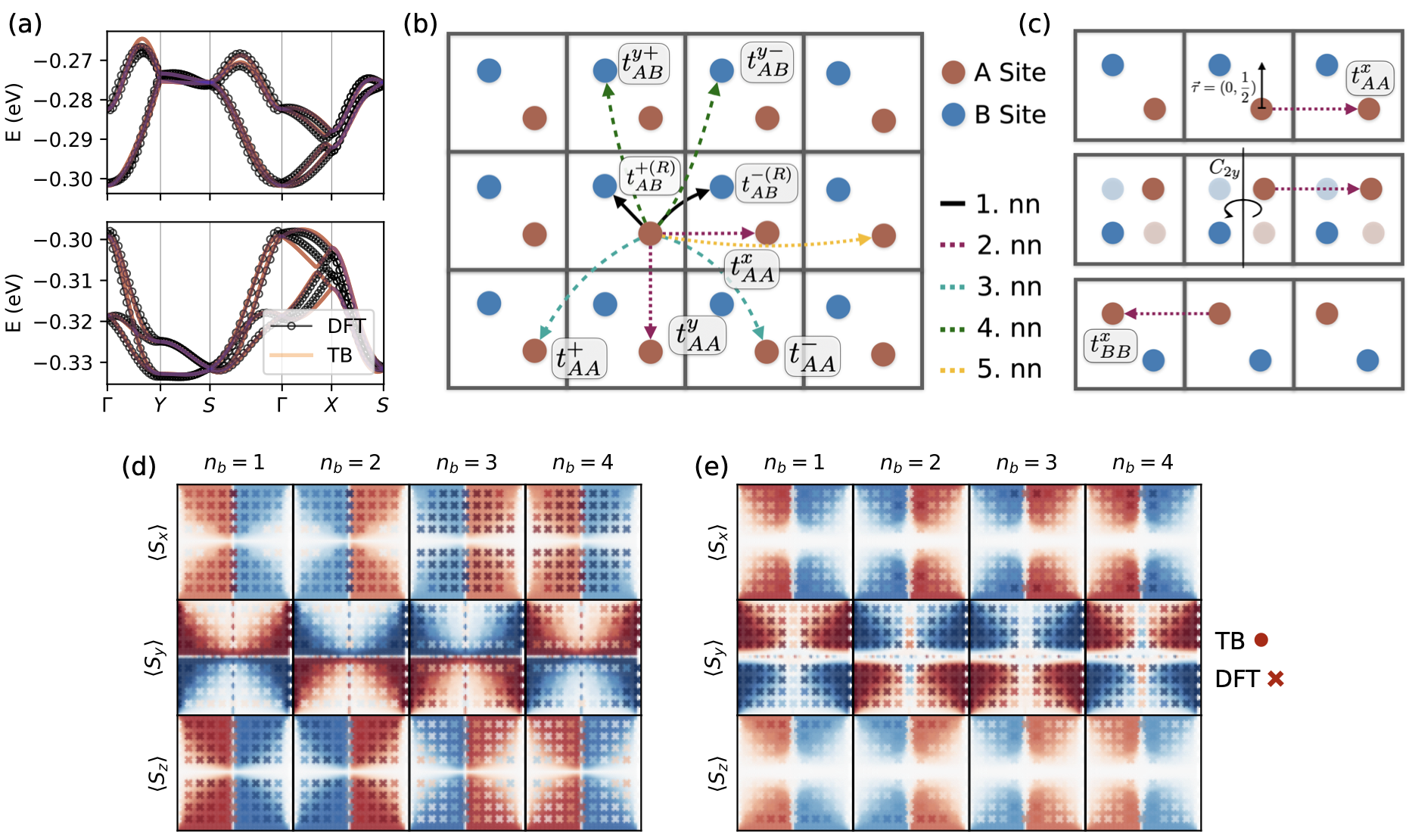}
    \caption{\textbf{Tight-binding model capturing SOC non-symmorphic symmetry and spin texture in twisted SnS.}
    (a) Comparison between the low-energy bandstructure obtained from DFT calculations and our 13-parameter TB fit. Band sets I/II are shown separately in the upper (lower) panel suggesting good agreement of the TB model with \textit{ab-initio} results. In particular, the hourglass crossing point is preserved indicating that the non-symmorphic symmetry $\widetilde{\mathcal{C}}_{2y}$ remains exact within our modeling approach. (b) Schematic sketch of the hopping amplitudes used in the 13-parameter tight-binding approach. Colors indicate the distance to the reference atom ($A$ site, orange sphere). The TB model takes up to 5th nearest-neighbor (nn) sites into account. (c) Schematic sketch of the impact of the non-symmorphic symmetry $\widetilde{\mathcal{C}}_{2y}$ on lattice sites and hopping amplitudes. As $\widetilde{\mathcal{C}}_{2y}$ consist of a shift by half a lattice vector in $y$ direction followed by a two-fold rotation around the $y$ axis (upper and middle panel), sublattices $A(B)$ are mapped onto each other. Therefore, hopping terms connecting same sublattices must be chosen identical to preserve the symmetry, i.e. $t_{AA}^x= t_{BB}^x$. (d), (e) Comparison of the spin texture obtained from DFT (cross) and TB (continuous color) calculations. Spin textures are plotted for different bands $n_b = 1 \dots 4$, where bands are labeled from the bottom to the top according to panel (a). The excellent agreement between DFT/TB model justifies the choice of nearest-neighbor Rahsba/Kane-Mele terms chosen in the manuscript.}
\label{fig:tb} 
\end{figure*}

To construct the effective low-energy tight-binding model that captures the topological features of the hourglass fermion it is essential to preserve the non-symmorphic symmetry $\widetilde{\mathcal{C}}_{2y}= i \tau_x \sigma_y$ and time-reversal symmetry $\mathcal{T} = -i \sigma_y \mathcal{K}$ of the underlying lattice. Here, the Pauli matrices $\sigma$ ($\tau$) act on the spin (sublattice) degree of freedom and $\mathcal{K}$ is the operator of complex-conjugation. In particular, $\widetilde{\mathcal{C}}_{2y}$ exchanges sublattices, i.e. $A \leftrightarrow B$, such that hopping amplitudes that connect same sublattices, i.e. two $A(B)$ sites with connecting vector $\bvec r_{AA(BB)}$, must be chosen identical to the hopping amplitude between two $B(A)$ sites if their connecting vectors are mapped onto each other under $\widetilde{\mathcal{C}}_{2y}$. This procedure is exemplified in Fig.~\ref{fig:tb} (c) showing the hopping amplitude between nearest-neighbor atoms in sublattice $A(B)$. To break the additional $\mathcal{C}_{2y}$ symmetry that is not present in the relaxed atomic structure of tSnS and to account for the anisotropy in $x/y$ that manifests due to the presence of the effective rectangular moiré unit cell, we allow for anisotropic hopping amplitudes to the right/left from each reference site as indicated by $+/-$ in Fig.~\ref{fig:tb} (b).

Furthermore, we model the strong intrinsic spin-orbit coupling of tSnS by Kane-Mele and Rashba SOC terms. The latter particularly break inversion and $SU(2)$ symmetry completely. The symmetry-allowed terms of our tight-binding fit are listed in Table~\ref{tab:tb}. To accurately capture the bandstructure and spin texture obtained from DFT, we set up a 13-parameter tight-binding model (the mass term $m_0$ may be absorbed into the chemical potential) including nearest-neighbor SOC terms and up to 5th nearest-neighbor kinetic energy hopping terms as sketched in Fig.~\ref{fig:tb} (b). 

\begin{table}[]
\begin{tabular}{c||ccccccccccccccc}
           & $m_0$   & $t_{AB}^+$ & $t_{AB}^-$ & $t_{AA}^x$ & $t_{AA}^y$ & $t_{AB}^{y+}$ & $t_{AB}^{y-}$ & $t_{AA}^{+}$ & $t_{AA}^{-}$ & $t_{AA}^{2x}$ & $t_{AA}^{2y}$ & $t_{AB}^{R+}$ & $t_{AB}^{R-}$ & $\phi_{AB}$ \\ \hline
Bandset I  & -281.60 & 3.97       & 3.06       & -0.10       & -4.02     & -1.25         & -0.89         & -1.26         & 0.85        & 0.00         & -0.71         & -0.56          & -0.45           & $\pi/30$  \\
Bandset II & -320.45 & 2.06       & 1.86       & 0.09       & 5.57      & 0.66          & 0.55          & -0.29        & -0.29        & 0.00          & 0.68          & -2.12          & -1.79         & $\pi/8$
\end{tabular}
\bigskip
\caption{Hopping parameters (in meV) for the 13-parameter tight-binding fit capturing the top valence band manifold (VBM) of twisted SnS. According to first-principle calculations, the 8 bands belonging to the VBM can be separated into two groups (Bandset I/II) with negligible residual coupling as suggested by the wave function profile in Fig.~\ref{fig:wavefunction}.}
\label{tab:tb}
\end{table}

The two-particle Hamiltonian consists of kinetic energy terms $t^{\text{kin}}$ that are spin-independent, i.e. $\propto \sigma_0$, Rashba SOC terms $t^{R}$ and Kane-Mele SOC terms causing $S_z$ polarization as indicated by the phase $e^{i \phi_{ij} s}$
\begin{equation}
H^0 = \sum_{ij s} t_{ij}^{\text{kin}}e^{i\phi_{ij}s} c^{\dagger}_{i s} c^{\phantom \dagger}_{js} + i \sum_{ij s s^{\prime}} t_{ij}^{R} c^{\dagger}_{i s} \left ( \bvec\sigma \times \bvec d_{ij} \right )_z c^{\phantom \dagger}_{j s^{\prime}} = H^{\text{kin}} + H^{\text{SOC}}.
\label{eq:tb_general}
\end{equation}
Here, $c^{(\dagger)}_{is}$ creates (annihilates) an electron with spin $s$ on moiré site $\bvec{r}_i$ that belongs to one of the sublattices $o \in \{A,B \}$. As the rectangular lattice possesses translational invariance, a description of the non-interacting theory is done most conveniently in Bloch momentum space. As the 4-band Hilbert space arises from $\hat{\bvec c}^{\dagger} = \{c^{\dagger}_{A \uparrow}, c^{\dagger}_{B \uparrow}, c^{\dagger}_{A \downarrow}, c^{\dagger}_{B \downarrow}\}$ this fixes our basis choice for the Bloch Hamiltonian. Defining the phase factors $\phi_{lm} := e^{-i\bvec k\cdot (l \bvec a_1 + \bvec a_2)}$ relative to the lattice vectors $\bvec a_{1(2)}$ of the rectangular lattice, the nearest-neighbor SOC terms of the Bloch Hamiltonian read
\begin{equation}
\begin{split}
H^{\text{SOC}}(\bvec k) = \tau_x & \left [ \frac{i}{2} t^{+R}_{AB} \left \{ \phi_{00}(-\sigma_x - \sigma_y) + \phi_{0\bar 1} (-\sigma_x + \sigma_y) \right \} \right . \\
&+ \frac{i}{2} t^{-R}_{AB} \left \{\phi_{\bar 10}(\sigma_x - \sigma_y) + \phi_{1\bar 1} (\sigma_x + \sigma_y) \right \}  \\
& + t_{AB}^+ \sigma^{\pm}(\phi_{AB}) \left \{ \phi_{00} + \phi_{0 \bar 1} \right \} + t_{AB}^- \sigma^{\pm}(-\phi_{AB}) \left \{ \phi_{10} + \phi_{1 \bar 1} \right \} \Bigg] + h.c.
\end{split}
\label{eq:tb_terms_soc}
\end{equation}
The appearance of the diagonal matrix $\sigma^{\pm}(\phi_{AB}) = \text{diag}(e^{i\phi_{AB}}, e^{-i\phi_{AB}})$ is reminiscent of the general Kane-Mele spin-orbit interaction, while the Rashba terms are proportional to $\sigma_{x(y)}$. We further use the notation $\bar l = -l$ for negative numbers. The vector connecting the sublattices $A(B)$ within the unit cell is given by $\bvec d_{AB}^{\text{intra}} = (-\frac{1}{2}, \frac{1}{2})$ and the three remaining vectors can be constructed by rotating by $\theta=\pi/2$. The kinetic energy terms that couple same sublattices read
\begin{equation}
\begin{split}
H^{\text{kin}}_{\tau_0}(\bvec k)= \sigma_0 &\tau_0  \left [ m_0 + t_{AA}^x \phi_{10} + t_{AA}^y \phi_{01} \right . \\
&+ \left. t_{AA}^{2x} \phi_{20} + t_{AA}^{2y} \phi_{02} \right ]\\
+ \sigma_0 &\tau_{AA}^{\pm}  \left [\phi_{1 1} + \phi_{\bar 1 1} \right ] + h.c.
\end{split}
\label{eq:tb_terms_kin_0}
\end{equation}
The appearance of the diagonal matrix $t_{AA}^{\pm} = \text{diag}(t_{AA}^+, t_{AA}^-)$ with nonequivalent hopping elements for sublattices $A(B)$ is a direct consequence of the non-symmorphic symmetry $\widetilde{\mathcal{C}}_{2y}$. The longer-ranged kinetic hopping elements connecting different sublattices read
\begin{equation}
H^{\text{kin}}_{\tau_x}(\bvec k) = \sigma_0 \tau_x \left [ t_{AB}^{y+}(\phi_{01}+\phi_{0 \bar 2})  + t_{AB}^{y-}(\phi_{11}+\phi_{1 \bar 2})  \right ].
\label{eq:tb_terms_kin_1}
\end{equation}
Altogether, the Hamiltonian for the band sets I/II is given by
\begin{equation}
H^0(\bvec k) = H^{\text{kin}}_{\tau_x}(\bvec k) + H^{\text{kin}}_{\tau_1}(\bvec k)  + H^{\text{SOC}}(\bvec k)
\label{eq:tb_total}
\end{equation}
The success of fitting the flat bands in twisted SnS within our low-energy tight-binding model is demonstrated in panel (a). By fitting the hopping parameters to the full ab-initio band structure for different angles, almost perfect agreement is achieved. In particular, the spin texture mandated by DFT calculations is captured accurately within our modeling approach, see Fig.~\ref{fig:tb} (d), (e) for band sets I/II. This shows that nearest-neighbor Rashba/Kane-Mele spin-orbit interactions are sufficient to reproduce the spin texture given by \textit{ab initio} calculations. A simple Python code to generate the Bloch Hamiltonian and the bandstructure of tSnS given the hopping parameters in table~\ref{tab:tb} is available online at \url{https://git.rwth-aachen.de/ammon.fischer/tight-binding-model-twisted-sns}

\section{Multi-orbital random-phase approximation}
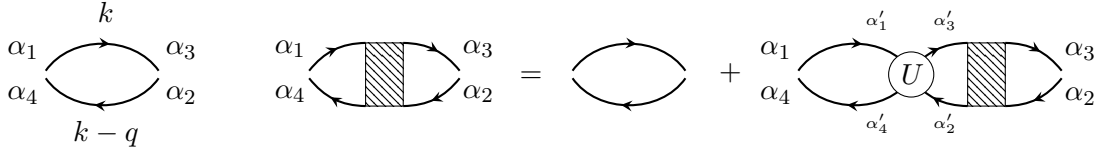
\begin{figure}[h]
    \centering
    \begin{tikzpicture}
	
	\def\gap{0.05}
	\def\hloop{0.42}
	
	\path [bend left=55, draw=black, line width=0.8pt, postaction={on each segment={mid arrow=black}}] 
	 (0,\gap)  to (1.5,\gap) node[below=-30, xshift = -0.7 cm]{$k$}
				      node[below=15, xshift = -0.7 cm]{$k-q$}
				      node[below=-15, xshift = -1.8 cm]{$\alpha_1$}
				      node[below=2, xshift = -1.8 cm]{$\alpha_4$}
				      node[below=-15, xshift = 0.3 cm]{$\alpha_3$}
				      node[below=2, xshift = 0.3 cm]{$\alpha_2$}
	(1.5,-\gap)    to  (0,-\gap)		      
	 ;

	\begin{scope}[shift={(3.5,0)}]

		\path [ bend left = 25, draw=black, postaction={on each segment={mid arrow=black}}, line width=0.8]
		(0, \gap) to (0.75, \hloop)				     
		(0.75, -\hloop) to (0, -\gap)
		(1.25, \hloop) to (2, \gap)
		(2, -\gap) to (1.25, -\hloop) 
							
				      node[below=-29, xshift = -1.5 cm]{$\alpha_1$}
				      node[below=-12, xshift = -1.5 cm]{$\alpha_4$}
				      node[below=-29, xshift = 1 cm]{$\alpha_3$}
				      node[below=-12, xshift = 1 cm]{$\alpha_2$}
		;

		\draw[pattern = north west lines, pattern color = black ]
		(0.75, -\hloop) rectangle ++(0.5, 2*\hloop) node[below=7, xshift = 1.7 cm]{$=$}
		;
								
	\end{scope}					
		
	\begin{scope}[shift={(7,0)}]

	\path [bend left=55, draw=black, line width=0.8pt, postaction={on each segment={mid arrow=black}}] 
	 (0,\gap)  to (1.5,\gap)
				      node[below=-7, xshift = 0.6 cm]{$+$}
	(1.5,-\gap)    to  (0,-\gap); 	
								
	\end{scope}

	\begin{scope}[shift={(10,0)}]
	\path [bend left=55, draw=black, line width=0.8pt, postaction={on each segment={mid arrow=black}}] 
	 (0,\gap)  to (1.5,\gap)
				      node[below=-15, xshift = -1.8 cm]{$\alpha_1$}
				      node[below=2, xshift = -1.8 cm]{$\alpha_4$}
	(1.5,-\gap)    to  (0,-\gap);

	\path [ bend left = 25, draw=black, postaction={on each segment={mid arrow=black}}, line width=0.8]
	(1.5, \gap) to (2.25, \hloop)
	(2.25, -\hloop) to (1.5, -\gap)
	(2.75, \hloop) to (3.5, \gap)
	(3.5, -\gap) to (2.75, -\hloop) 
				      node[below=-28, xshift = 1 cm]{$\alpha_3$}
				      node[below=-11, xshift = 1 cm]{$\alpha_2$}

				      node[below=-40, xshift = -1.7 cm]{\tiny $\alpha_1^{\prime}$}
				      node[below=-40, xshift = -0.8 cm]{\tiny $\alpha_3^{\prime}$}
				      node[below=-1, xshift = -1.7 cm]{\tiny $\alpha_4^{\prime}$}
				      node[below=-1, xshift = -0.8 cm]{\tiny $\alpha_2^{\prime}$}

	;

	\draw[pattern = north west lines, pattern color = black ]
	(2.25, -\hloop) rectangle ++(0.5, 2*\hloop) 
	;

	\fill[white] (1.5,0) circle (0.3);
	\draw (1.5,0) circle (0.3) node[xshift = 0.02 cm]{$U$};	
			
	\end{scope}

\end{tikzpicture}
    \caption{Left: Loop diagram for the non-interacting 4-point correlation function (susceptibility) $\chi_{\alpha_1 \dots \alpha_4}^{PH}$ defined in Eq.~\eqref{eq:generalized_loop_ph}. The solid lines denote fermionic propagators with frequency and momentum dependence $k=(ik_0, \bvec k)$. The labels on the in-going and out-going legs $\alpha_1 \dots \alpha_4$ are ordered in standard notation and refer to the electrons' quantum numbers $\alpha = (o,s)$ in terms of orbital (sublattice) $o \in \{A, B\}$ and spin $s$ degree of freedom. Right: Diagrammatic representation of the RPA enhanced correlation function (susceptibility) as defined in Eq.~\ref{eq:generalized_loop_ph}. The emerging Dyson equation denotes a summation to infinite order in the interaction vertex $U_{\alpha_1 \dots \alpha_4}$. Internal indices connecting interaction vertex and the (non-interacting) susceptibilities are summed over implicitly and are denoted with small labels.}
    \label{fig:rpa_loop}
\end{figure}

In this section, we summarize the random-phase approximation (RPA) formalism applied in the main text that captures the collective spin-/charge excitations (DW) and superconducting instabilities in twisted SnS. In the absence of spin rotational $SU(2)$ symmetry caused by the strong Rashba-type spin-orbit coupling (SOC) in the material, the electrons' spin remains no longer a well-defined quantum number when transforming the multi-orbital Bloch Hamiltonian 
\begin{equation}
H^0_{\alpha, \alpha^{\prime}}(\bvec k) = u_{\alpha b}^{\phantom \dagger}(\bvec k)e_b(\bvec k) u_{\alpha^{\prime} b}^{\dagger}(\bvec k)
\label{eq:o2b}
\end{equation}
to band space $(\alpha, b)$, where $\alpha = (o, s)$ denotes the orbital (sublattice)- and spin degrees of freedom. It is therefore that spin- and charge excitation generally mix in the absence of $SU(2)$ symmetry for arbitrary momentum transfer $\bvec q$ and a description beyond the decoupling in longitudinal and transverse fluctuations sufficient in the case of $SU(2)$ symmetry is needed~\cite{romer2021superconducting,xiong2021spin,graser2009near,wu2021nature,scherer2018spin-orbit}. Before discussing details of the weak-coupling expansion as given by the RPA, we may define the free Matsubara Green's function in orbital-momentum (frequency) space that connects different interactions vertices appearing in the diagrammatic expansion
\begin{equation}
g_{\alpha,\alpha^{\prime}} (i\omega, \bvec k) = \left [i\omega - (H^0(\bvec k ))  )\right ]_{\alpha, \alpha^{\prime}}^{-1} = \sum_{b} u_{\alpha b}(\bvec k) g_b (i \omega, \bvec k) u_{b \alpha^{\prime}}^{*}(\bvec k) =
 	\begin{tikzpicture}[baseline = -2.5]
	
		\path [draw=black,postaction={on each segment={mid arrow=black}}, line width=0.8pt]
		(0,0) -- (2,0) 	node[above=2, xshift = -2 cm]{$\alpha$}
					node[above=2, xshift = -0. cm]{$\alpha^{\prime}$} 
					node[below=2, xshift = -1 cm]{$(b, \bvec k)$}
		;
	\end{tikzpicture}.
\label{green}
\end{equation}
Here, $u_{\alpha b}$ are the orbital-to-band transformations that render the free Green's function $g_b(i \omega, \bvec k) = (i \omega -e_b(\bvec k))^{-1}$ diagonal. \\ \\
It is the core of the random-phase approximation as single-channel approximation of the effective two-particle interaction vertex $\Gamma^{(4)}$ to expand the two-particle Green's function $\chi^{\text{RPA}}(\bvec q)$ in a series of bubble- and ladder diagrams to infinite order in $U$ such to account for all possible one-loop corrections as shown in Fig.~\ref{fig:rpa_loop}. By rewriting the renormalized interactions in the particle-hole (ph) and particle-particle (pp) channel, tendencies towards either spin-/charge density wave order (DW) or superconducting order (SC) are resolved. To this end, it is beneficial to introduce the generalized (undressed) four-point correlation function $\chi^{PH/PP}(\bvec q)$ in the particle-hole and particle-particle channel

\begin{equation}
\chi^{PH}_{\alpha_1 ..\alpha_4}(q) = -\frac{1}{N} \sum_{\bvec k} \sum_{b, b^{\prime}} \left[ M ^{PH} (\bvec k, \bvec q)\right]_{\alpha_1 ..\alpha_4}^{b, b^{\prime}} \frac{n_F(\epsilon_b(\bvec k)) - n_F(\epsilon_{b^{\prime}}(\bvec k - \bvec q))}{i q_0 + \epsilon_b(\bvec k) - \epsilon_{(b^{\prime}}(\bvec k - \bvec q)}
\label{eq:generalized_loop_ph}
\end{equation}
\begin{equation}
\chi^{PP}_{\alpha_1 ..\alpha_4}(q) = -\frac{1}{N} \sum_{\bvec k} \sum_{b, b^{\prime}} \left[ M^{PP} (\bvec k, \bvec q)\right]_{\alpha_1 ..\alpha_4}^{b, b^{\prime}} \frac{n_F(\epsilon_b(\bvec k)) - n_F(1 - \epsilon_{b^{\prime}}(-\bvec k - \bvec q))}{i q_0 + \epsilon_b(\bvec k) + \epsilon_{(b^{\prime}}(-\bvec k - \bvec q)}.
\label{eq:generalized_loop_pp}
\end{equation}
Here, we adopt standard notation and label in-going legs ($\alpha_1, \alpha_2$) and out-going ($\alpha_3, \alpha_4$) legs accordingly. The tensor $M^{PH/PP} (\bvec k, \bvec q)$ roots in the contributions of the unitary orbital-to-band transformations Eq.~\eqref{eq:o2b} and are given by
\begin{equation}
\left[ M^{PH} (\bvec k, \bvec q)\right]_{\alpha_1 ..\alpha_4}^{b, b^{\prime}} = u_{\alpha_1 b}(\bvec k) u^{*}_{\alpha_3 b}(\bvec k) u_{\alpha_2 b^{\prime}}(\bvec k - \bvec q) u^{*}_{\alpha_4 b^{\prime}}(\bvec k - \bvec q)
\label{eq:m_ph}
\end{equation}
\begin{equation}
\left[ M^{PP} (\bvec k, \bvec q)\right]_{\alpha_1 ..\alpha_4}^{b, b^{\prime}} = u_{\alpha_1 b}(\bvec k) u^{*}_{\alpha_3 b}(\bvec k) u_{\alpha_2 b^{\prime}}(-\bvec k - \bvec q) u^{*}_{\alpha_4 b^{\prime}}(-\bvec k - \bvec q).
\label{eq:m_pp}
\end{equation}
In numerical simulations it is favorable to achieve the highest possible momentum resolution of the Fermi surface in order to capture nesting effects that will determine the spin-/charge response of the system. Therefore, we resort to a refinement procedure that relies on averaging the response function for each momentum point $\bvec k$ appearing in the sum of Eq.~\eqref{eq:generalized_loop_ph_RPA} over a set of finer momentum points $\{\bvec{k}_f \}_{\bvec k}$
\begin{equation}
\chi^{PH}_{\alpha_1 ..\alpha_4}(q) = -\frac{1}{N} \sum_{\bvec k} \sum_{\{\bvec k_f\}_{\bvec k}} \sum_{b, b^{\prime}} \left[ M ^{PH} (\bvec k_f, \bvec q)\right]_{\alpha_1 ..\alpha_4}^{b, b^{\prime}} \frac{n_F(\epsilon_b(\bvec k_f)) - n_F(\epsilon_{b^{\prime}}(\bvec k_f - \bvec q))}{i q_0 + \epsilon_b(\bvec k_f) - \epsilon_{(b^{\prime}}(\bvec k_f - \bvec q)}.
\label{eq:generalized_loop_ph_fine}
\end{equation}
In practice, this is achieved by discretizing the 2D Brillouin zone with an equidistant momentum mesh with $N_{\bvec k} \times N_{\bvec k}$ coarse momentum points and averaging over a fine momentum mesh with $N_{\bvec k_f} \times N_{\bvec k_f}$ point such that the electronic dispersion is effectively resolved for $(N_{\bvec k}N_{\bvec k_f})^2$ points. For the phase diagram presented in Fig. 3 (a) in the manuscript, we used $N_{\bvec k}=48$ and $N_{\bvec k_f} = 40$. \\

In terms of the generalized four-point susceptibility, the expansion of the two-particle Greens function $\chi^{\text{RPA}}$ in bubble and ladder diagrams is straight forward by using Hugenholtz diagrams representing symmetrized interaction vertices, see Fig.~\ref{fig:rpa_loop}. In the presence of only an repulsive Hubbard interaction, i.e.
\begin{equation}
U_{\alpha_1 \dots \alpha_4} =
\begin{cases}
    +U, & \text{if } o_1=o_2=o_3=o_4;\, s_1=s_3,\, s_2=s_4 \\
    -U, & \text{if } o_1=o_2=o_3=o_4;\, s_1=s_2,\, s_3=s_4 \\
    0, & \text{otherwise}
\end{cases}
\label{eq:U_vertex}
\end{equation}
we can resum this series to infinite order in $U$ to obtain the full RPA corrected two-particle Greens function $\overline{\chi^{PH}} = \chi^{\text{RPA}}$
\begin{equation}
\begin{split}
 \overline{\chi^{PH}}_{\alpha_1 ..\alpha_4}(\bvec q) &= \chi^{PH}_{\alpha_1 ..\alpha_4}(\bvec q) + \chi^{PH}_{\alpha_1, \alpha_4^{\prime}, \alpha_1^{\prime}, \alpha_4}(\bvec q) U_{\alpha_1^{\prime},\alpha_2^{\prime},\alpha_3^{\prime}, \alpha_4^{\prime}}(\bvec q) \overline{\chi^{PH}}_{\alpha_3^{\prime},\alpha_2,\alpha_3,\alpha_2^{\prime}}(\bvec q).
\end{split}
\label{eq:generalized_loop_ph_RPA_}
\end{equation}
Here, we only focus on the static part of the effective interaction vertex, i.e. $iq_0 = 0$. Einstein notation applies as double indices are summed over implicitly. To solve this linear equation for $\overline{\chi^{PH}}$, it is beneficial to re-order the legs of each rank-4 tensor according to the new positions $(1,4,3,2)$ to express the tensor sums in Eq.~\eqref{eq:generalized_loop_ph_RPA_} as batched matrix-matrix multiplication in $\bvec q$. This yields
\begin{equation}
\begin{split}
 \overline{\chi^{PH}}_{(\alpha_1, \alpha_4)(\alpha_3, \alpha_2)}(\bvec q) &= \chi^{PH}_{(\alpha_1, \alpha_4)(\alpha_3, \alpha_2)}(\bvec q) + \chi^{PH}_{(\alpha_1, \alpha_4)(\alpha_1^{\prime}, \alpha_4^{\prime})}(\bvec q) U_{(\alpha_1^{\prime},\alpha_4^{\prime})(\alpha_3^{\prime}, \alpha_2^{\prime})}(\bvec q) \overline{\chi^{PH}}_{(\alpha_3^{\prime},\alpha_2^{\prime})(\alpha_3,\alpha_2)}(\bvec q) \\
&= \left [ \matrix{\chi}^{PH} + \matrix{\chi}^{PH} \cdot \matrix{U} \cdot \overline{\matrix{\chi}^{PH}} \right ]_{(\alpha_1, \alpha_4)(\alpha_3, \alpha_2)}(\bvec q) \\
&= \left [\frac{\mathds{1}}{\mathds{1}-\matrix{\chi}^{PH} \cdot \matrix{U}} \cdot \matrix{\chi}^{PH} \right ]_{(\alpha_1, \alpha_4)(\alpha_3, \alpha_2)}(\bvec q) 
\end{split}
\label{eq:generalized_loop_ph_RPA}
\end{equation}
To retain the standard notation, the legs need to be re-ordered accordingly. From the expression above it is clear that (i) spin- and charge correlations are generally mixed in the absence of $SU(2)$ symmetry and (ii) that the RPA corrected susceptibility encounters divergences when the denominator becomes singular. This 'generalized' Stoner criterion for multi-orbital systems can be used to estimate the critical interaction strength $U_c$ that marks the onset of spin-/charge density wave (DW) order in the system. Divergences occur first at the interaction value for which the maximal eigenvalue $\lambda_{\text{max}}$ of the (non-hermitian) product $\matrix{U} \cdot \matrix{\chi}^{PH}$ becomes $U_c = 1/\lambda_{\text{max}}$. This procedure is used to separate the phase diagram as function of filling $\nu$ and interaction strength $U$ into paramagnetic and DW ordered regions.

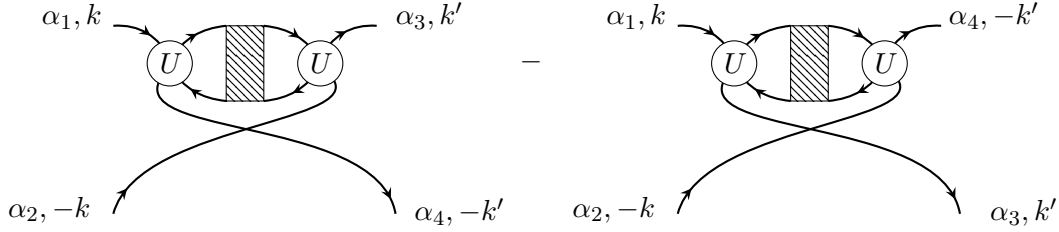
\begin{figure}[h]
    \centering
     	\begin{tikzpicture}
	
	\def\gap{0.05}
	\def\hloop{0.5}

	\path [bend left=35, draw=black, line width=0.8pt, postaction={on each segment={mid arrow=black}}] 
	 (0.75,\hloop)  to (1.5,\gap) node[below=-25, xshift = -1.3 cm]{$\alpha_1, k$}
	 ; 

	\path [in=100, out=-135, draw=black, line width=0.8pt, postaction={on each segment={past arrow=black}}] 
	(1.5,-\gap)    to  (4.5,-4*\hloop) 	 node[below=-10, xshift = 0.85 cm]{$\alpha_4, -k^{\prime}$};	

	\path [out=75, in=-35, draw=black, line width=0.8pt, postaction={on each segment={pre arrow=black}}] 
	 (0.75,-4*\hloop) to (3.5,-\gap) node[below=45, xshift = -3.6 cm]{$\alpha_2, -k$};

	\path [ bend left = 25, draw=black, postaction={on each segment={mid arrow=black}}, line width=0.8]
	(1.5, \gap) to (2.25, \hloop)
	(2.25, -\hloop) to (1.5, -\gap)
	(2.75, \hloop) to (3.5, \gap)
	(3.5, -\gap) to (2.75, -\hloop) 				      

	;

	\draw[pattern = north west lines, pattern color = black ]
	(2.25, -\hloop) rectangle ++(0.5, 2*\hloop) 
	;

	\path [bend left=35, draw=black, line width=0.8pt, postaction={on each segment={mid arrow=black}}] 
	 (3.5,\gap)  to (4.25,\hloop) node[below=-13, xshift = 0.7 cm]{$\alpha_3, k^{\prime}$} ;

	\fill[white] (1.5,0) circle (0.3);
	\draw (1.5,0) circle (0.3) node[xshift = 0.02 cm]{$U$}
	;	

	\fill[white] (3.5,0) circle (0.3);
	\draw (3.5,0) circle (0.3) node[xshift = 0.02 cm]{$U$}
	node[below=-8, xshift = 2.8 cm]{$-$}
	;

	\begin{scope}[shift={(7.5,0)}]

	\path [bend left=35, draw=black, line width=0.8pt, postaction={on each segment={mid arrow=black}}] 
	 (0.75,\hloop)  to (1.5,\gap) node[below=-25, xshift = -1.3 cm]{$\alpha_1, k$}
	 ; 

	\path [in=100, out=-135, draw=black, line width=0.8pt, postaction={on each segment={past arrow=black}}] 
	(1.5,-\gap)    to  (4.5,-4*\hloop) 	 node[below=-10, xshift = 0.85 cm]{$\alpha_3, k^{\prime}$};	

	\path [out=75, in=-35, draw=black, line width=0.8pt, postaction={on each segment={pre arrow=black}}] 
	 (0.75,-4*\hloop) to (3.5,-\gap) node[below=45, xshift = -3.6 cm]{$\alpha_2, -k$};

	\path [ bend left = 25, draw=black, postaction={on each segment={mid arrow=black}}, line width=0.8]
	(1.5, \gap) to (2.25, \hloop)
	(2.25, -\hloop) to (1.5, -\gap)
	(2.75, \hloop) to (3.5, \gap)
	(3.5, -\gap) to (2.75, -\hloop) 				      

	;

	\draw[pattern = north west lines, pattern color = black ]
	(2.25, -\hloop) rectangle ++(0.5, 2*\hloop) 
	;

	\path [bend left=35, draw=black, line width=0.8pt, postaction={on each segment={mid arrow=black}}] 
	 (3.5,\gap)  to (4.25,\hloop) node[below=-13, xshift = 0.7 cm]{$\alpha_4, -k^{\prime}$} ;

	\fill[white] (1.5,0) circle (0.3);
	\draw (1.5,0) circle (0.3) node[xshift = 0.02 cm]{$U$}
	;	

	\fill[white] (3.5,0) circle (0.3);
	\draw (3.5,0) circle (0.3) node[xshift = 0.02 cm]{$U$}
	;

	\end{scope}

	\end{tikzpicture}
    \caption{Diagrams contributing to the pairing vertex within the fluctuation-exchange approximation. The vertex must is anti-symmetrized to fulfill fermionic anti-commutation relations, which can be achieved by swapping the out-going legs $3 \to 4$, starting from the simple RPA expansion.}
    \label{fig:flex}
\end{figure}
For interaction values $U<U_c$, the effective coupling in the ph-channel did not diverge yet and fluctuations around DW ordered phases may provide the pairing glue for unconventional superconducting states in the system. To capture these contributions from fluctuation-exchange~\cite{romer2021superconducting}, we rewrite the effective interaction in the pp-channel $\Gamma^{\text{PP}}(\bvec k, \bvec k^{\prime})$ to further track possible pairing instabilities. The diagrammatic contributions shown in Fig.~\ref{fig:flex} translate to 
\begin{equation}
\begin{split}
\Gamma^{\text{PP}}(\bvec k, \bvec k^{\prime},\alpha_1 .. \alpha_4) &= U_{\alpha_1,\alpha_1^{\prime},\alpha_4^{\prime}, \alpha_4}(\bvec q) \overline{\chi}^{PH}_{\alpha_1^{\prime},\alpha_3^{\prime},\alpha_2^{\prime},\alpha_4^{\prime}}(\bvec q) U_{\alpha_3^{\prime},\alpha_3, \alpha_2, \alpha_2^{\prime}}(\bvec q) \delta_{\bvec q, \bvec k - \bvec k^{\prime}} \\
&\overset{\mathclap{}}{=}  U_{(\alpha_1,\alpha_4)(\alpha_1^{\prime},\alpha_4^{\prime})}(\bvec q) \overline{\chi}^{PH}_{(\alpha_1^{\prime},\alpha_4^{\prime})(\alpha_3^{\prime},\alpha_2^{\prime})}(\bvec q) U_{(\alpha_3^{\prime},\alpha_2^{\prime})(\alpha_3, \alpha_2)}(\bvec q) \delta_{\bvec q, \bvec k - \bvec k^{\prime}} \\
&= \left [ \matrix{U} \cdot \matrix{\overline{\chi}^{PH}} \cdot \matrix{U}\right ]_{(\alpha_1, \alpha_4)(\alpha_3, \alpha_2)}(\bvec k - \bvec k^{\prime}).
\end{split}
\label{eq:generalized_flex_vertex}
\end{equation}
From the first to the second line, we transposed each rank-4 tensor according to the order $(1,4,2,3)$ to rewrite the tensor product as batched matrix-matrix multiplication in $\bvec q$.  To account for all contributions from bubble and ladder diagrams, we anti-symmetrize the pairing vertex by swapping the outgoing legs $3 \leftrightarrow 4$ such that we finally arrive at
\begin{equation}
\Gamma^{\text{PP}}_{\sigma} = \frac{1}{2} \left [\Gamma^{\text{PP}}(\bvec k, \bvec k^{\prime},\alpha_1,\alpha_2,\alpha_3,\alpha_4) - \Gamma^{\text{PP}}(\bvec k, - \bvec k^{\prime},\alpha_1,\alpha_2,\alpha_4,\alpha_3)  \right ].
\label{eq:generalized_flex_vertex_symmetrized}
\end{equation}

\section{Analyzing spin-/charge density wave and superconducting order}

This section aims to establish the connection between (i) the dressed two-particle Greens function $\chi^{\text{RPA}}(\bvec q)$ and the spin-/charge correlation functions in the presence of broken $SU(2)$ symmetry and (ii) the effective particle-particle interaction vertex $\Gamma^{\text{PP}}$ and emerging superconducting instabilities and their symmetry.

First, to analyze spin- and charge excitations, we resort to the spin-spin (charge-charge) correlation function in imaginary time
\begin{equation}
\begin{split}
\chi_{ij}(i\omega, \bvec q) &= \int_{0}^{\beta} d \tau e^{i \omega \tau}\langle \mathrm{T}_{\tau} X_{i}(\bvec q, \tau) X_{j}(-\bvec q, 0) \rangle,
\end{split}
\label{eq:corr_functions_imag}
\end{equation}
where $X_i(\bvec q, \tau)=N(\bvec q, \tau)$ if $i=0$ describes the density-density response of the system and $X_i(\bvec q, \tau)=S(\bvec q, \tau)$ if $i \in \{x,y,z\}$ describes the spin-spin correlations associated to the spatial directions of the lattice structure. Furthermore, $\tau \in [0, \beta]$ denotes the imaginary time argument with inverse temperature $\beta=1/T$ and $T_{\tau}$ is the time-ordering operator. In this context, the spin (charge) operators are defined as
\begin{equation}
\begin{split}
S_i(\bvec q, \tau) &= \frac{1}{N} \sum_{k} \sum_{o,s,s^{\prime}} c^{\dagger}_{o,s}(\bvec k, \tau) \frac{\left [\sigma_i \right ]_{s s^{\prime}}}{2} c^{\phantom \dagger}_{o,s^{\prime}}(\bvec k + \bvec q, \tau) \\
N(\bvec q, \tau) &= \frac{1}{N} \sum_{k} \sum_{o,s,s^{\prime}} c^{\dagger}_{o,s}(\bvec k, \tau) \left [\sigma_0 \right ]_{s s^{\prime}} c^{\phantom \dagger}_{o,s^{\prime}}(\bvec k + \bvec q, \tau), 
\end{split}
\end{equation}
where $\sigma_i$ denotes the $i$-th Pauli matrix and $\sigma_0$ refers to the identity matrix. Comparing the expression of the spin-/charge correlation function Eq.~\eqref{eq:corr_functions_imag} with the dressed four-point susceptibility Eq.~\eqref{eq:generalized_loop_ph_RPA_} being an four-fermion expectation value shows that the correlation functions can be recovered by forming appropriate linear combinations of the entries of $\chi^{\text{RPA}}_{\alpha_1 \dots \alpha_4}(\bvec q)$. The correct terms are recovered by contraction with Pauli matrices on both sides
\begin{equation}
\label{eq:}
\chi_{ij}(\bvec q) = \sum_{o_1 \dots o_4} \sum_{s_1 \dots s_4} \left [\sigma_i \right ]_{s_1 s_4} \chi^{\text{RPA}}_{\substack{s1 \dots s_4 \\ o_1 \dots o_4}}(\bvec q) \left [\sigma_j \right ]_{s_3 s_2}, 
\end{equation}
where we traced over the orbital degrees of freedom. This is sufficient when analyzing in which channel the leading instability occurs. The leading channel is identified by the maximum value of any $|\chi^{\text{RPA}}(\bvec q)|$ at given wave vector $q$. In general, divergencies in multiple channel can occur and the resulting DW instability will be a mixture of spin-/charge density waves in the case of broken $SU(2)$ symmetry as mentioned in the beginning. \\ \\

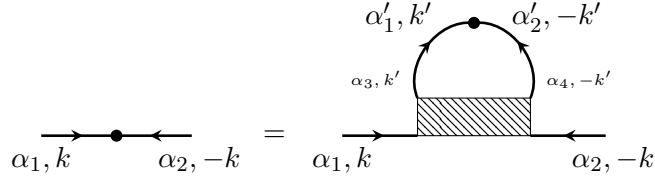
\begin{figure}[h]
    \centering
     	\begin{tikzpicture}

	\def\lvertex{1.5}
	\def\hvertex{0.5}
	\def\hloop{1.5}
	\def\loopbend{60}
	\def\lw{1.}

	\path [draw=black,postaction={on each segment={mid arrow=black}}, line width= \lw pt]
	(0,0) -- (1,0) 	node[below=0, xshift = -1 cm]{$\alpha_1, k$}
	(2,0) --  (1,0)	node[below=0, xshift = +1.1 cm]{$\alpha_2, -k$}
				node[xshift = +2.1 cm]{$=$}
	;
	
	\fill[black] (1,0) circle (0.08);

	\begin{scope}[shift={(4,0)}]
	
	\draw[pattern = north west lines, pattern color = black ]
	(1, 0) rectangle ++(\lvertex, \hvertex) 
	;
	
	\path [draw=black,postaction={on each segment={mid arrow=black}}, line width=\lw pt]
	(0,0) -- (1,0) 					node[below=0, xshift = -1 cm]{$\alpha_1, k$}
	(2 + \lvertex, 0) --  (1 + \lvertex, 0)	node[below=0, xshift = +1.1 cm]{$\alpha_2, -k$}
	; 
	
	\path [bend left=\loopbend, draw=black, line width=\lw pt, postaction={on each segment={mid arrow=black}}] 
	 (1,\hvertex)  to (1+ 0.5*\lvertex, \hloop) 	node[below=-12, xshift = -1.0 cm]{$\alpha_1^{\prime}, k^{\prime}$}
	 								node[below=14, xshift = -1.3 cm]{\tiny $\alpha_3, k^{\prime}$}
	 ;
	
	\path [bend right=\loopbend, draw=black, line width=\lw pt, postaction={on each segment={mid arrow=black}}] 
	 (1 + \lvertex,\hvertex)  to (1+ 0.5*\lvertex, \hloop) 	node[below=-12, xshift = 1.1 cm]{$\alpha_2^{\prime}, -k^{\prime}$}
											node[below=14, xshift = 1.4 cm]{\tiny $\alpha_4, -k^{\prime}$}
	 ;
	
	\fill[black] (1 + 0.5*\lvertex, \hloop) circle (0.08);
	
	\end{scope}

	\end{tikzpicture}
    \caption{Linearized gap equation for the superconducting order parameter $\Delta_{\alpha_1 \alpha_2}(\bvec k)$. The hatched area denotes the effective vertex in the particle-particle channel $\Gamma_{\sigma}^{\text{PP}}$ as defined in Eq.~\eqref{eq:generalized_flex_vertex_symmetrized}}
    \label{fig:lingap}
\end{figure}

Next, we turn to the discussion of the superconducting ordering tendencies. Given the effective interaction vertex in the particle-particle channel $\Gamma^{\text{PP}}$ Eq.~\eqref{eq:generalized_flex_vertex_symmetrized}, in the following we assume $\Gamma^{\text{PP}} = \Gamma^{\text{PP}}_{\sigma}$ to be the symmetrized interaction vertex implicitly, the symmetry of the leading superconducting instabilities is further analyzed by solving the linearized gap equation in the vicinity to the critical temperature $T_c$ of the superconducting phase. For multi-orbital systems, the linearized gap equation as diagrammatically shown in Fig.~\ref{fig:lingap}  takes the form of a generalized (non-hermitian) eigenvalue problem
\begin{equation}
\begin{split}
\Delta_{\alpha_1 \alpha_2}(\bvec k) &= \frac{1}{\beta N} \sum_{k^{\prime}} \sum_{\alpha_1^{\prime} \alpha_2^{\prime}} \sum_{\alpha_3 \alpha_4} \Gamma^{\text{PP}}_{\alpha_1, \alpha_2, \alpha_3, \alpha_4}(\bvec k, \bvec k^{\prime}) G_{\alpha_3 \alpha_1^{\prime}}(k^{\prime}) G_{\alpha_4 \alpha_2^{\prime}}(-k^{\prime}) \Delta_{\alpha_1^{\prime} \alpha_2^{\prime}}(\bvec k^{\prime}) \\
& = \frac{1}{N} \sum_{\bvec k^{\prime}} \sum_{\alpha_1^{\prime} \alpha_2^{\prime}} \sum_{\alpha_3 \alpha_4} \Gamma^{\text{PP}}_{\alpha_1, \alpha_2, \alpha_3, \alpha_4}(\bvec k, \bvec k^{\prime}) \chi^{\text{PP}}_{\alpha_3 \alpha_4 \alpha_1^{\prime} \alpha_2^{\prime}}(\bvec k^{\prime}, \bvec q = 0) \Delta_{\alpha_1^{\prime} \alpha_2^{\prime}}(\bvec k^{\prime}) \\
&= \frac{1}{N} \sum_{\bvec k^{\prime}} \sum_{\alpha_1^{\prime} \alpha_2^{\prime}} \Pi^{\text{PP}}_{(\bvec k, \alpha_1, \alpha_2) (\bvec k^{\prime}, \alpha_1^{\prime}, \alpha_2^{\prime})} \Delta_{(\bvec k^{\prime}, \alpha_1^{\prime} \alpha_2^{\prime})} \\
&= \lambda_{\text{SC}} \left [ \hat\Pi^{\text{PP}} \hat\Delta \right ]_{(\bvec k, \alpha_1, \alpha_2)}.
\end{split} 
\label{eq:vertex_pp_loop}
\end{equation}
From the first to the second line, we identified the particle-particle loop Eq.~\eqref{eq:generalized_loop_pp} as the product of the two (free) Greens functions with momentum dependence $(k, -k)$. As for the effective interaction we only consider zero-momentum Cooper pairs, i.e. $\bvec q = 0$ in the particle-particle loop, as these will give dominant contributions among all Fermi liquid instabilities. From the second to the third line, we contracted the effective pairing vertex $\Gamma^{\text{PP}}(\bvec k, \bvec k^{\prime})$ and the particle-particle loop $\chi^{\text{PP}}(\bvec k^{\prime})$ by executing the sum over ($\alpha_3, \alpha_4$) for each momentum $\bvec k^{\prime}$. By reordering indices as indicated in the third line underlines the character of the linearized gap equation of being an effective (non-hermitian) eigenvalue problem in orbital/spin-momentum space. The leading eigenvalue $\lambda_{\text{SC}}$ indicates the superconducting solution with highest transition temperature $\lambda_{\text{SC}} \propto T_c$ and the corresponding eigenfunction $\Delta_{\alpha_1 \alpha_2}(\bvec k)$ transforms as one of the irreducible representations of the systems symmetry groups and hence yields the symmetry of the underlying SC order parameter. \\
\\ 
The disadvantage of solving the linearized gap equation in orbital space lies in the numerical cost associated with the solution of the non-hermitian eigenvalue problem imposed by Eq.~\eqref{eq:vertex_pp_loop}. The joint index $(\bvec k, \alpha_1, \alpha_2)$ indicating the dimension of the effective matrix $\hat\Pi^{\text{PP}}$ scales as $N_b^2 \times N_{\bvec k}$. For the $N_b = 4$ bands in tSnS as well as $N_{\bvec k}=48\times 48$ being the number of (coarse) momentum points used to resolve the superconducting order parameter in Fig. 4 of the manuscript, the resulting matrix has dimension $\approx 37'000 \times 37'000$.  Thus, this approach is limited to few band models as diagonalizing even larger non-hermitian matrices is merely feasible. \\
On the other hand, the advantage of solving the linearized gap equation in orbital space lies in its gauge invariance, which is crucial for determining topological properties of symmetry broken phases as all information about topology is encoded in the phase of the Bloch states. The reason for it is that the solution of the linearized gap in orbital-space Eq.~\eqref{eq:vertex_pp_loop} only contains the particle-particle loop $\chi^{\text{PP}}$, which is gauge invariant as the orbital-to-band transformations Eq.~\eqref{eq:o2b} only appear in complex-conjugated pairs and thus possible gauge phases cancel out. In principle, only scattering of electrons in the immediate vicinity of the Fermi surface will contribute notably to Fermi liquid instabilities with non-vanishing order parameter $\Delta(\bvec k)$. Therefore, most approaches in literature~\cite{graser2009near,romer2021superconducting} aim to first project the effective interaction $\Gamma^{\text{PP}}(\bvec k, \bvec k^{\prime}) \to \Gamma^{\text{PP}}_{b_1 \dots b_4}(\bvec k_F, \bvec k^{\prime}_F)$ to band space such that the momenta and corresponding bands are restricted to the Fermi surface sheets at each respective filling. However, such a transformation sensitively depends on the specific gauge of the orbital-to-band transformations $u_{\alpha b}(\bvec k)e^{i\phi_b(\bvec k)}$ that in general will carry an arbitrary gauge phase $\phi_b(\bvec k)$ due to numerical diagonalization of the Bloch Hamiltonian Eq.~\eqref{eq:o2b} at each momentum point $\bvec k$. Transforming the effective particle-particle vertex $\Gamma^{\text{PP}}$ to band space the projected vertex must hence be written as

\begin{equation}
\begin{split}
\Gamma^{\text{PP}}_{b_1 \dots b_4}(\bvec k_F, \bvec k_F^{\prime}) = &\sum_{\alpha_1 \dots \alpha_4} u_{\alpha_1 b_1}^*(\bvec k_F) u_{\alpha_2 b_2}^*(-\bvec k_F) \Gamma^{\text{PP}}_{\alpha_1 \alpha_2 \alpha_3 \alpha_4}(\bvec k_F, \bvec k^{\prime}_F) u_{\alpha_3 b_3}^{\phantom *}(\bvec k^{\prime}_F) u_{\alpha_4 b_4}^{\phantom *}(-\bvec k^{\prime}_F) \\
&\times e^{i(\phi_{b_3}(\bvec k_F^{\prime})+ \phi_{b_4}(-\bvec k_F^{\prime}) - \phi_{b_1}(\bvec k_F) - \phi_{b_2}(-\bvec k_F))},
\end{split}
\label{eq:transformation_vertex_pp}
\end{equation}
and is therefore not gauge invariant. In the presence of time-reversal symmetry $\mathcal{T}$ and restricting the analysis to intraband pairing only, i.e. $b_1=b_2$ and $b_3 = b_4$, this can be cured by explicitly constructing the time-reversal partners of the orbital-to-band transformations $\mathcal{T}{|u_{\bvec k} \rangle} = i \sigma_y \tau_0 u^*_{os}(-\bvec k)$. This way, the phase difference $e^{i(\phi_{b}(\bvec k_F^{\prime}) -  \phi_{b}(\bvec k_F^{\prime}))} = 1$ cancels out and gauge invariance is recovered. Nevertheless, this still imposes issues at points in momentum space with degenerate energies, e.g. the Weyl nodal point in tSnS, as the gauge in the degenerate subspace is not uniquely defined and furthermore interband processes become relevant, such that the approximate intraband solution may not be reliable any more. Furthermore, in the presence of arbitrary spin-orbit interactions, the electrons spin is no longer a well defined quantum number as it mixes with the orbital degrees of freedom. This makes it impossible to classify the superconducting order parameter with respect to spin-singlet/triplet or orbital character. This is only possible in certain situation, where one may define a smooth pseudo-spin basis such that certain bands carry well-defined spin expectation value~\cite{romer2021superconducting}, which is not the case for tSnS.

\section{Hourglass Superconductivity in twisted SnS}

The eigenfunctions $\Delta_{\alpha_1, \alpha_2}(\bvec k)$ obtained by solving the linearized gap equation~Eq.\eqref{eq:vertex_pp_loop} can be classified according to their transformation behaviour under the non-symmorphic symmetry group $\widetilde{\mathcal{C}}_{2y}=\{\mathcal{C}_{2y}|0\frac{1}{2}\}$. As the latter inherits from the point group symmetry $\mathcal{C}_{2y}$, solutions can be labeled to whether they transform even/odd under $\widetilde{C}_{2y}$, i.e. $U^{\dagger}(\bvec k) \Delta (\bvec k) U(\bvec k) = \pm \Delta (\bvec k)$ with the unitary matrix $U(\bvec k) = i \tau_x \sigma_y e^{ik_y/2} \mathcal{R}_{k_x \to -k_x}$. This is because the action of the translation by half a Bravais vector in $y$-direction $\tau = (0, \frac{1}{2})$ is included in the phase factor $e^{i \bvec k \cdot \bvec \tau} = e^{i k_y/2}$, which cancels in the expression $U^{\dagger}(\bvec k) \Delta (\bvec k) U(\bvec k)$. Using $\mathcal{R}^{\dagger}_{k_x \to -k_x} \Delta(k_x, k_y) \mathcal{R}_{k_x \to -k_x} = \Delta(-k_x, k_y)$, we may further simplify the expression to $U^{\dagger}(\bvec k) \Delta (\bvec k) U(\bvec k) = \widetilde{U}^{\dagger}(\bvec k) \Delta_{\alpha_1, \alpha_2}(-k_x, k_y) \widetilde{U}(\bvec k)$. The matrix $\widetilde{U}_{\alpha_1, \alpha_2}(\bvec k)$ in the basis choice $\hat{\bvec c}^{\dagger} = \{c^{\dagger}_{A \uparrow}, c^{\dagger}_{B \uparrow}, c^{\dagger}_{A \downarrow}, c^{\dagger}_{B \downarrow}\}$ then reads

\begin{equation} \widetilde{U}(\bvec k) = 
\begin{pmatrix}
0 & 0 & 0 & \phi_{\bar10}(\bar{\bvec k})\\
0 & 0 & \phi_{\bar1 \bar 1}(\bar{\bvec k}) & 0\\
0 & -\phi_{\bar10}(\bar{\bvec k}) & 0 & 0\\
-\phi_{\bar1 \bar 1}(\bar{\bvec k}) & 0 & 0 & 0
\end{pmatrix}
\label{eq:U_matrix}
\end{equation}
where the phase factors $\phi(\bar{\bvec k})$ need to be introduced in order to ensure numerical equivalence and $\bar{\bvec k} = (-k_x, k_y)$ is the $\mathcal{C}_{2y}$ transformed momentum. The reason for it is that by applying $\mathcal{C}_{2y}$ to entries of the Bloch Hamiltonian or SC order parameter combining different sublattices, e.g. $\Delta_{AB}^{\uparrow \downarrow}(\bvec k)$, bonds that lie originally in the same unit cell (and therefore carry no phase $\phi(\bar{\bvec k})$ resembling their momentum dependence) are mapped to different unit cells under $\mathcal{C}_{2y}$. Therefore, they pick up a phase dependence that needs to be compensated when numerically comparing to non-transformed order parameter. \\ \\

\begin{figure*}
    \includegraphics[width=0.8\linewidth]{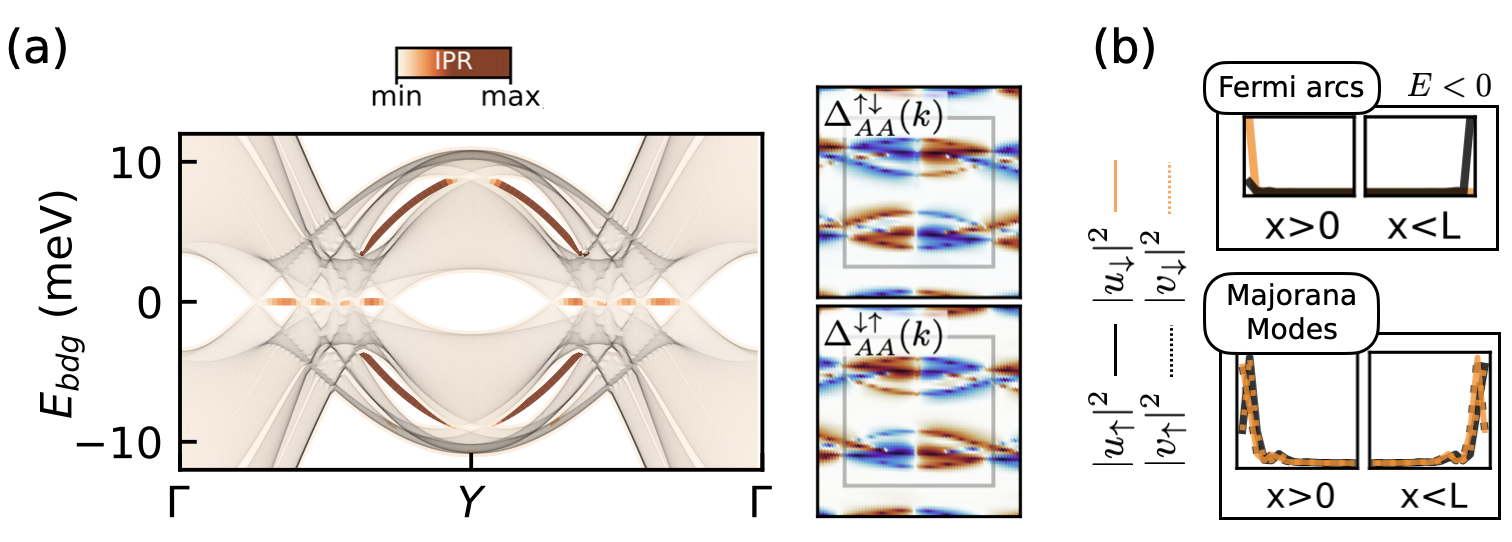}
    \caption{\textbf{Sub-leading superconducting solution $\Delta_u$ appearing in the phase diagram of tSnS.} (a) Bogoliubov de-Gennes (BdG) quasi-particle bandstructure of a slab geometry with open boundaries in $x$ direction. The sub-leading SC instability preserves the Fermi arcs and features a Majorana zero energy mode that is localized at the edge. The localization is visualized by the inverse participation ratio (IPR) that is color-coded for each band in (a). The corresponding wave function profiles are shown in panel (b). Due to particle-hole symmetry, amplitudes of particle- and holes with different spin orientations are identical, i.e. $|u_{\uparrow}|^2 = |v_{\downarrow}|^2$. 
}
\label{fig:majorana_fermion} 
\end{figure*}

To further analyze the quasiparticle properties of the SC instability and determine the existence of potential localized edge modes, we study the order parameter within mean-field theory. The resulting mean-field Hamiltonian can be rewritten in the  Nambu spinor basis $\psi^{\dagger}_{\bvec{k}} = ( \vec{c}^{ \, \dagger}_{\bvec{k}\uparrow} \,\,\, \vec{c}^{\, \dagger}_{\bvec{k}\downarrow} \,\,\,  \vec{c}^{\phantom \dagger}_{-\bvec{k}\uparrow} \,\,\,  \vec{c}^{\phantom \dagger}_{-\bvec{k}\downarrow}   )^{\mathrm{T}}$. The full $2N$-dimensional Hamiltonian is of Bogoliubov-de Gennes (BdG) form 
\begin{equation}
\mathsf{H}_{\text{MF}} = \sum_{\bvec{k}} \psi^{\dagger}_{\bvec{k}}  
\begin{pmatrix} \hat{H}_0(\bvec{k}) & \hat{\Delta}(\bvec{k}) \\ \hat{\Delta}^{\dagger}(\bvec{k}) & -\hat{H}_{0}^*(-\bvec{k})  \\\end{pmatrix}
\psi^{\phantom{\dagger}}_{\bvec{k}} = \sum_{\bvec{k}} \left ( \hat{U}_{\bvec{k}} \psi_{\bvec{k}}\right )^{\dagger} \begin{pmatrix} \hat{E}_{\bvec{k}} &0 \\ 0 & -\hat{E}_{\bvec{k}}  \end{pmatrix} \left ( \hat{U}_{\bvec{k}} \psi_{\bvec{k}}\right )^{\phantom \dagger} \\
\hat{U}^{\phantom \dagger}_{\bvec{k}},
\label{bdg}
\end{equation}
and is diagonalized by a block-structured unitary transform $\hat{U}_{\bvec{k}}$ 
\begin{equation}
\hat{U}^{\phantom \dagger}_{\bvec{k}} = \begin{pmatrix} \hat{u}_{\bvec{k}} & -\hat{v}_{\bvec{k}} \\ \hat{v}^*_{\bvec{k}}& \hat{u}_{\bvec{k}}  \\\end{pmatrix} \quad \text{and} \quad \hat{U}^{\dagger}_{\bvec{k}} \hat{U}^{\phantom \dagger}_{\bvec{k}} = \mathds{1}.
\label{bdg_unitary}
\end{equation}
The matrices $ \hat{u}_{\bvec{k}} \left (  \hat{v}_{\bvec{k}} \right )$ are $N_b \times N_b$ matrices, which describe the particle- (hole) amplitudes of the fermionic Bogoliubov quasiparticles $\gamma_{\bvec{k}}$ with energies $\pm E_{\bvec{k}}$. The particle-hole symmetric BdG quasi-particle spectrum for the sub-leading instability $\Delta_u$ is shown in Fig.~\ref{fig:majorana_fermion} (a), whereas the leading instability $\Delta_g$ is presented in the manuscript, see Fig. 4. The inverse participation ration (IPR) in the superconducting phase $\text{IPR}(\bvec k, b) = \sum_{o,s} |u_{os,b}(\bvec k)|^4 + |v^*_{os, b}(\bvec k)|^4$ indicates strong localization of (i) Fermi arcs that are preserved in the sub-leading solution due to the nodal line of the order parameter $\Delta_u$ at the position of the Weyl nodal point at $\bvec k = Y/2$ and (ii) zero-energy Majorana modes that appear in the BdG quasi-particle spectrum. Analyzing the wave function profiles of both topological modes demonstrates that these are highly localized at the systems edge, see Fig.~\ref{fig:majorana_fermion} (b).  While the edge modes have different spin polarization in case of the Fermi arcs, i.e. $|u_{\uparrow}|^2$ and $|u_{\downarrow}|^2$ are enhanced at opposite boundaries, the Majorana modes are characterized by identical amplitudes of particle- and holes with different spin orientations due to the present of particle-hole symmetry, i.e. $|u_{\uparrow}|^2 = |v_{\downarrow}|^2$ at $E=0$.

\bibliography{References.bib}